\documentclass[useAMS,usenatbib]{mn2e}

\usepackage{amssymb}
\usepackage{graphicx}
\DeclareGraphicsExtensions{.jpg,.pdf,.png,.eps,.pdf}
\usepackage{url}
\usepackage{color}
\usepackage{tabulary}
\usepackage{multirow}
\usepackage{amsmath}
\usepackage{amsfonts}
\usepackage{ulem}

\graphicspath{ {pictures/} }
\newcommand{\beq}{\begin{equation}}
\newcommand{\eeq}{\end{equation}}
\newcommand{\simlt}{\mathrel{\hbox{\rlap{\hbox{\lower4pt\hbox{$\sim$}}}\hbox{$<$}}}}
\newcommand{\simgt}{\mathrel{\hbox{\rlap{\hbox{\lower4pt\hbox{$\sim$}}}\hbox{$>$}}}}

\newcommand{\s}{\;\mathrm{s}}
\newcommand{\dd}{\partial}
\newcommand{\Msol}{\;\mathrm{M}_{\odot}}

\newcommand{\pc}{\;\mathrm{pc}}
\newcommand{\yr}{\;\mathrm{yr}}
\newcommand{\Myr}{\;\mathrm{Myr}}

\def\apjl{ApJL}
\def\apj{ApJ}
\def\mnras{M.N.R.A.S.}
\def\aap{A\&A}
\def\nat{Nat.}
\def\araa{Ann. Rev. A\&A}

\def\aj{AJ}

\def\aapr{A\&A Rev.}

\defcitealias{Inayoshi+16}{IHO16}

\title[IMBHs from Pop~III remnants]{Intermediate-mass black holes from Population III remnants in the first galactic nuclei}

\author[T. Ryu et al.]{Taeho Ryu$^{1}$\thanks{email: taeho.ryu@stonybrook.edu}, Takamitsu L. Tanaka$^{1}$, Rosalba Perna$^{1,2}$, Zolt\'an Haiman$^{3}$\\ 
$^{1}$Department of Physics and Astronomy, Stony Brook University, Stony Brook, NY 11794-3800, USA\\
$^{2}$Fellow Adjoint of JILA, 440 UCB, Boulder, CO 80309-0440, USA\\
$^{3}$Department of Astronomy, Columbia University, 550 W. 120th Street, New York, NY 10027, USA}

\begin{document}

\maketitle

\label{firstpage}

\begin{abstract}
We report the formation of intermediate-mass black holes (IMBHs) in
suites of numerical $N$-body simulations of Population III remnant
black holes (BHs) embedded in gas-rich protogalaxies at redshifts
$z\ga 10$.  We model the effects of gas drag on the BHs' orbits, and
allow BHs to grow via gas accretion, including a mode of
hyper-Eddington accretion in which photon trapping and rapid gas
inflow suppress any negative radiative feedback.  Most initial BH
configurations lead to the formation of one (but never more than one)
IMBH in the center of the protogalaxy, reaching a mass of
$10^{3-5}\Msol$ through hyper-Eddington growth.  Our results suggest a
viable pathway to forming the earliest massive BHs in the centers of
early galaxies.  We also find that the nuclear IMBH typically captures
a stellar-mass BH companion, making these systems observable in
gravitational waves as extreme mass-ratio inspirals (EMRIs) with
\textit{eLISA}.
\end{abstract}


\section{Introduction}
\label{sec:intro}

Virtually every nearby massive galaxy harbors a supermassive black
hole (SMBH) in its nucleus \citep{KormendyHo13,Magorrian+96}. Tight correlations
between SMBHs and the properties of their host galaxies, as well as
the phenomenology of quasars and luminous active galactic nuclei,
suggest that SMBHs play a role in shaping their environment both on
local \citep{Silk98,Fabian12} and cosmological \citep{Madau+04,
  RicottiOstriker04, TOP16} scales.

How SMBHs form is one of the most fundamental open problems in
astrophysics.  The observation of quasars powered by $\sim 10^9\Msol$
SMBHs at $z\approx 6-7$, less than a Gyr after the Big Bang
\citep{Fan+01, Willott+03, Willott+10, Mortlock+11, Venemans+13},
places strong constraints on theoretical models for their origin (see
reviews by, e.g. \citealt{Volonteri10,Haiman13}).  The two most often
discussed hypotheses are that they grew from either (i) the stellar BH
remnants of the first generation of $\sim 100 \Msol$ stars (Population
III, hereafter Pop~III, stars; \citealt{HaimanLoeb01,
  MadauRees01,YM04, VR06, dyformula1}), or (ii) the remnants of the
``direct collapse'' of $>10^5\Msol$ gas clouds that avoided
fragmentation into Pop~III stars \citep{Oh02, BrommLoeb03,
  Koushiappas+04, VolonteriRees05, Shapiro05, Begelman+06, SS06,
  LodatoNatarajan06, WA08, RH09a, SSG10, Shang+10,
  Latif+13,TanakaLi14}.  Both scenarios require that some of the first
SMBHs grew at a (logarithmically averaged) rate $\dot{M}\sim 10 L_{\rm
  Edd}/c^2$ \citep[e.g.][]{Tanaka14}, where $L_{\rm Edd}\propto M$ is
the Eddington luminosity for a BH of mass $M$, and $c$ is the speed of
light.  This value is comparable to the accretion rate producing the
Eddington luminosity with the radiative efficiency $\eta\equiv
L/(\dot{M}c^2)\sim 0.1$ expected in thin discs.\footnote{Radiatively
  efficient discs are theoretically expected to have $\eta \sim
  0.1-0.4$ \citep[e.g.][]{Shapiro05}, in agreement with the mean value
  $\sim 0.07 - 0.1$ inferred from the So\l tan-Paczynski argument,
  comparing the mass density of SMBHs with the quasar luminosity
  density (e.g. \citealt{MerloniHeinz08, Shankar+10}, but see
  \citealt{Shankar+2016} for a recent argument for a higher value).}

In this paper, we examine the possibility that the first nuclear SMBHs
originated from \textit{hyper-Eddington} accretion onto Pop~III
remnants---i.e. from a growth mode where $\dot{M}\gg \dot{M}_{\rm
  Edd}\equiv L_{\rm Edd}/c^2$.  This is motivated by theoretical
models of optically thick accretion flows in which photons are trapped
and advected inside the accretion flow. In such ``radiatively
inefficient'' accretion modes the luminosity and radiation feedback of
the flow are quenched, allowing accretion rates much higher than those
corresponding to the Eddington limit for radiatively efficient discs
(see \citealt{Begelman1979} for spherical flows and \citealt{Abramowicz+98} for slim 
accretion discs).  Several studies have investigated whether such
an accretion mode contributed to the growth of the first SMBHs
\citep{VolonteriRees05,Begelman12,Madau+14,AlexanderNatarajan2014,
  Volonteri+15,PacucciFerrara2015, Pacucci+2015,Lupi+16,Pezzulli+16}.

We focus our attention on the recent work by \citet[][hereafter
  IHO16]{Inayoshi+16}, who found hyper-Eddington accretion solutions
in spherically symmetric radiation-hydrodynamics simulations of
Bondi-like accretion. They found, in broad agreement with previous
simulations \citep{Milos+09,ParkRicotti12}, that radiative feedback
typically limits the accretion rate to values comparable or below
$\dot{M}_{\rm Edd}$.  This radiative feedback arises from
photoionization and heating of the gas near the Bondi radius, and
occurs even for flows that are highly optically thick to electron
scattering, and for which trapping of the radiation limits the
luminosity emerging from the photosphere below $L_{\rm Edd}$. However,
\citetalias{Inayoshi+16} also found that for sufficiently large
ambient gas densities, the combination of the large ram pressure of
the inflowing gas and photon trapping inhibit radiative feedback.  The
accretion flow in this regime is steady and unimpeded from the Bondi
rate; following \citetalias{Inayoshi+16} we refer to this as
hyper-Eddington accretion.

We perform $N$-body simulations of Pop~III remnant BHs in a model
protogalactic distribution of gas and dark matter (DM) at $z\ga 10$,
subjecting the BHs to dynamical friction, and allowing them to accrete
in the manner found by \citetalias{Inayoshi+16}.  The goal of our
simulations is to follow the coupled growth and orbital evolution of a
small cluster of stellar-remnant BHs, and to evaluate whether such BHs
can reasonably be expected to reach the hyper-Eddington regime, and
grow rapidly into more massive BHs.

In our models, we find that Pop~III BHs indeed frequently grow into
intermediate-mass black holes (IMBHs) with masses over $10^3\Msol$,
and even into supermassive holes (SMBHs) with masses $10^5\Msol$.  We
further find that these I/SMBHs always form after a lower-mass BH has
eroded its orbit, and settled near the center of the model
protogalaxy. This suggests that hyper-Eddington accretion is a viable
mechanism for forming nuclear SMBHs in early galaxies.

In addition, we report that our simulations always produce only a
single I/SMBH.  This is because once these massive BHs dominate the
central potential, subsequent BHs dragged into the dense central
regions reach high velocities that prohibit their growth. Instead,
they are captured as stellar-mass BHs in a bound orbit.  This suggests
that I/SMBH formation is typically accompanied by so-called
extreme-mass ratio inspirals (EMRIs), the merger of compact objects
with mass ratios $\ll 1$ that are one of the main targets of the
planned space-borne gravitational-wave detector \textit{eLISA}.

The rest of this paper is organized as follows.  In \S2, we describe
the setup of our simulations, including the properties of the Pop~III
remnant BHs and of their host galaxy, as well as the numerical schemes
used to simulate their orbital evolution and growth.  We present and
discuss our main results in \S3. Several implications and theoretical
caveats are discussed further in \S4.  We summarize our conclusions in
\S5.

\section{Numerical model}\label{sec:numericalsetup}

In this section, we provide an overview of our simulations. We first
describe the properties of the parent galaxy, and the initial
conditions for the small cluster of stellar-mass BHs.  We then
describe the equations of motion we solve to follow the growth of BHs
and their interactions and dynamics, as well as the numerical scheme
we use to solve these equations.  A key feature of our model is a
prescription that allows rapid growth by gas accretion, based on the
recent numerical study of Bondi-like hyper-Eddington accretion with
radiation by \citetalias{Inayoshi+16}.

\subsection{Protogalaxy + BH population model}
\label{subsec:model}

\subsubsection{Protogalactic gas cloud and DM halo}

We consider a small cluster of Pop~III-remnant BHs embedded in a
protogalactic, so-called atomic-cooling DM halo, with a
virial temperature of $T_{\rm vir}\geq 10^{4} {\rm K}$ and mass $\ga
10^{7-8}\Msol$ at $z\sim 15-20$.  Such a system is a plausible outcome
of lower-mass Pop~III-forming haloes growing either via accretion and
minor mergers, or via major mergers.  Massive stars are expected to
form earlier, in the lower-mass progenitor ``minihaloes'', and leave
behind stellar-mass BH remnants.  However, the UV radiation and/or
supernova (SN) explosions of the progenitor star of such a BH can
unbind the gas from the shallow potential well of its host
minihalo. The remnant BHs are then expected to be starved
(e.g. \citealt{AWA09}), and not surrounded again by dense gas and grow
until they are incorporated into more massive atomic-cooling haloes.

The gas collapsing inside an atomic cooling halo is expected to cool
and condense, and to develop a nearly isothermal density profile with
${\rm d}\ln \rho / {\rm d}\ln r \approx -2$
\citep{Oh02,Wise07,Regan09,Shang+10, Latif+14,Regan+14}.  Under the
assumption that the metallicity in this protogalaxy remains low, and
${\rm H_2}$ cooling is disabled, the temperature will remain near the
HI atomic cooling floor of $8000~{\rm K}$.  This configuration is
expected to be rare, as it requires a large UV (Lyman-Werner) flux
from a bright neighbour, forming near-simultaneously
\citep{Dijkstra+08,Agarwal+12,Visbal+14}.  In the presence of metal
and/or ${\rm H_2}$-cooling, the large-scale inflow rate is expected to
be slower, as a result of the lower sound speed,
  reducing the normalization of the density profile (see,
  e.g. \citealt{Shang+10}).

Our fiducial protogalaxy model is based on the metal- and ${\rm
  H_2}$--free atomic cooling halo, and consists of a DM component with
a \cite{Navarro+97} (NFW) density profile, and an isothermal gas
profile that behaves as $\propto r^{-2}$ at large radii.  In order to
avoid a mathematical singularity at the origin, we introduce a core
region of size $r_{\rm c}$ inside which the density is nearly
constant.

The matter distribution in our model protogalaxy can be summarized as follows:
\begin{align}\label{eq:rhogas}
\rho_{\rm bg}(r)&= \rho_{\rm gas}(r)+\rho_{\rm NFW}(r) \\
\rho_{\rm gas}(r)&=  \frac{\rho_{\rm c}}{1+(\frac{r}{r_{\rm c}})^{2}},
\end{align}
where $\rho_{\rm NFW}(r)$ is the NFW profile with concentration
parameter $C=9$ and the virial radius is $r_{\rm vir}\simeq 1{\rm~kpc}$. 
The virial radius is defined as the radius within which the average matter density 
is 180 times the cosmological critical density.
The values of $r_{\rm c}$ and $n_{\rm c}$ are
determined by normalizing the gas profile to satisfy the cosmological
ratio of baryon to DM mass inside $r_{\rm vir}$. We
take $r_{\rm c}=0.003\pc$ and $n_{\rm c}=2.5\times 10^{10}{\rm
  cm}^{-3}$. These values are consistent with those found in the
highest-resolution adaptive mesh refinement simulations to date
\citep{Regan+14}.
For numerical convenience, we slightly modify the NFW profile (which
scales as $r^{-1}$ at small radii) by requiring that the DM density
does not exceed the gas density for $r\leq r_{\rm NFW,c}$, a radius at which $\rho_{\rm gas}=\rho_{\rm NFW}$.
This modification only affects the region inside $\sim 10^{-3}r_{\rm
  c}$, and does not appreciably affect our results. The density
profile of DM and gas remains fixed and unchanging throughout our simulations;
the possible impact of this large simplification is addressed below.

\begin{table}
	\centering
	\setlength\extrarowheight{5pt}
	\begin{tabulary}{1\linewidth}{l l}
		\hline
		Redshift & $z\approx 15-20$\\
		Sound speed & $c_{\rm s}\approx 10~{\rm km}\s^{-1}$\\
		Halo mass & $M_{\rm vir, halo}\approx 5\times 10^{7}-10^{8}\Msol$\\
		Halo virial temperature & $T_{\rm vir} \gtrsim10000~{\rm K}$\\
		Initial radial distance of BHs & $r_{i}(t=0)\le 100\pc$\\
		Initial separations between BHs & $r_{ij} (t=0)\ge 10 \pc$\\		
		Initial speed of BHs & $v_{i}(t=0) \le c_{\rm s}$\\
		Mean molecular weight & $\mu_{\rm m}=1$ \\
		Gas density profile & $\rho_{\rm gas}\propto r^{-2}$ (isothermal sphere) \\
		Dark matter density profile & $\rho_{\rm DM}$=NFW \\
		Concentration parameter	 &	C=9\\
		Virial radius	&	$r_{\rm vir}\simeq 1{\rm~kpc}$\\
		Core gas density & $n_{\rm c}=2.5\times 10^{10}~{\rm cm^{-3}}$\\
		Core radius for gas &  $r_{\rm c}=3\times 10^{-3}\pc$\\
		Core radius for DM & $r_{\rm NFW,c}\approx10^{-6}\pc$\\
		Initial stellar mass function range & $25\Msol \leq M_{\star}\leq 140 \Msol$\\
		Initial stellar mass function slope & $\alpha=0.17$\\
                Total run time & 		$t_{\rm run}=500~{\rm Myrs}$	\\
		\hline
	\end{tabulary}
\caption{Choices of the values of different physical parameters
  defining our protogalaxy + BH cluster model. See text for details. 
Note that the italic subscripts \textit{i} and \textit{j} are indices representing the BHs.}
\label{tab1}
\end{table}

\subsubsection{Pop~III-remnant BHs}
\label{subsubsec:BHsetup}

Within our spherically symmetric halo, we place a small cluster of ten
BHs inside a radius of $100\pc$. This represents $\approx 10\%$ of
the virial radius of a halo just above the atomic cooling threshold, and
is intended to correspond to the spatial extent of a star-forming
region, or the region over which the BHs are initially spread after
merging events.  Each hemisphere is divided by the polar angle into
five compartments of equal shape and size, and a single BH is placed
at a randomly chosen radius and angular position inside each
compartment (i.e., ten compartments, and one BH per compartment).

We also require that the initial distance between each pair of BHs is
larger than $10\pc$.  Each BH is given a random initial velocity, so
that its speed is no larger than the gas sound speed $c_{\rm s}\approx
10\,{\rm km}\,{\rm s}^{-1}$ and the radial component of its velocity
is nonpositive (i.e. it is not flying away from the origin; this is
intended to mimic the outcome of recent mergers).

The masses of the Pop III stars as the BH progenitors are assigned
randomly from an initial mass function (IMF)
$\frac{dN}{dM_{\star}}=M_{\star}^{-\alpha}$ with $\alpha=0.17$ \citep{stacy13}.  We
adopt a minimum mass of $M_{\text{min},\star} =25\Msol$, and a maximum mass
of $M_{\text{max},\star}=140\Msol$. The latter value is motivated by the
fact that more massive stars are expected to result in
pair-instability supernovae and leave no BH remnant \citep{Heger+03}.
While stars more massive than $\approx 260\Msol$ may form BHs, such
large masses may be precluded by UV feedback in the protostellar
stages \citep[e.g.][]{Hosokawa+11}.

After the stellar masses are drawn, they are converted to BH masses $M$
using the following fitting formulae provided by \cite{Tanaka+12},
which are based on simulations by \cite{Zhang+08}:
\begin{equation}
M=
\begin{cases}
\frac{3}{4}(M_{\star}-20\Msol)+2\Msol & $\text{if }$ M_{\star}\leq
45\Msol\\
\frac{5}{12}(M_{\star}-20\Msol)& $\text{if }$ M_{\star} > 45\Msol\,.
\end{cases}
\end{equation}
The average mass of a Pop~III star is $\langle M_{\star}\rangle
\approx 80\Msol$, and that of a remnant BH is $\langle M\rangle
\approx 25\Msol$.  All simulations are run until either a tightly
bound BH pair with a semimajor axis $a\le 1~{\rm pc}$ forms, or
until a physical time of $t_{\rm run}=500\Myr$ has elapsed---whichever
occurs first.\\

\begin{figure}
\centering
\includegraphics[width=1.0\linewidth]{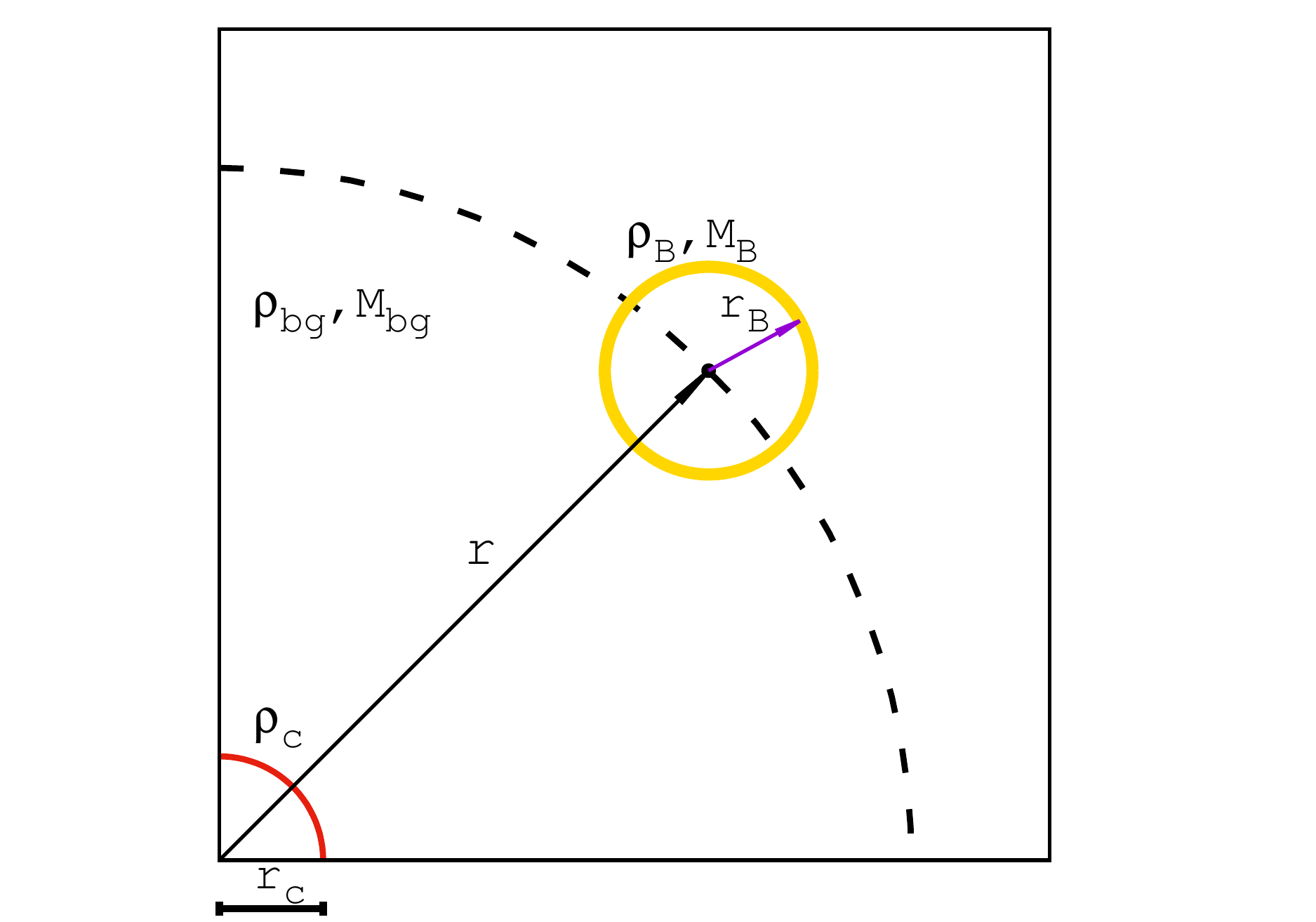}
\caption{Schematic representation of one of the BHs in the halo,
  defining our notation.  $M_{\rm bg}(<r)$ denotes the mass of the
  background material (gas + DM) enclosed within the radial position
  $r$ of the BH, while $M_{\rm B}$ represents the mass of gas enclosed within
  the Bondi radius of the BH. }
\label{fig:configuration}
\end{figure}

The properties of the host halo and the BHs adopted in our simulations
are summarized in Table \ref{tab1}, and a schematic diagram of the
halo + BH system is illustrated in Figure~\ref{fig:configuration}.
The radial coordinate of the BH is denoted by $r$.  The gas density
profile is spherically symmetric, and has a flat central core of size
$r_{\rm c}$.  The sum of the gas and DM densities define the total
background matter density $\rho_{\rm bg}$.  We denote the mass
enclosed inside the instantaneous radial coordinate of the BH by
$M_{\rm bg}(<r)$.  The Bondi radius $r_{\rm B}$ (defined in
\S\ref{subsec:mdot} below) defines a spherical region around each BH
that we call the ``Bondi sphere.''  We will make use of the average
density around the surface of this sphere, $\langle \rho\rangle_{\rm
  B}$, and the total mass enclosed inside the Bondi sphere, $M_{\rm
  B}$.

\subsection{The equations of motion}
\label{subsec:eom}

We use an $N$-body code to integrate the equations of motion and mass
growth for each BH embedded in the model protogalaxy.  The motion of
the BHs is determined by the following forces: (i) their mutual
gravitational attraction (including post-Newtonian terms up to 2.5th
order), (ii) dynamical friction/drag from the surrounding medium, and
(iii) the gravitational potential of the background matter.  We also
account for (iv) the deceleration due to mass growth via accretion.
The resulting equation of motion for the $i^{\rm th}$ BH includes the
sum of the five forces:
\beq
\textbf{a}_{i}=\textbf{a}_{{\rm N},i}+\textbf{a}_{{\rm PN},i}+\textbf{a}_{{\rm df},i}+\textbf{a}_{{\rm bg},i}+\textbf{a}_{{\rm acc},i}
\label{eq:eom}
\eeq
This equation is iteratively integrated at every time step, and used
to update the positions and velocities with the $N$-body algorithm
described in \S~\ref{subsec:code} below. We next describe each
contribution in detail.

\begin{enumerate}
	
\item \noindent\textit{Gravitational attraction between BHs}\\
This contribution is dominated by the standard Newtonian expression, 

\begin{equation}
\label{eq:Newton-pairs}
\textbf{a}_{{\rm N}, i} = -\sum_{j\neq i} G~M_{{\rm BH},j}~
\frac {\dd ~\Phi(r_{ij})}{\dd ~r_{ij}}~
\frac{\textbf{r}_{i}-\textbf{r}_{j}}{r_{ij}},
\end{equation}
where $G$ is the gravitational constant, $\Phi$ is the pairwise
gravitational potential, $\textbf{r}_{i}$ is the displacement of the
$i^{{\rm th}}$ BH from the center of the host DM halo, and
$r_{ij}\equiv |\textbf{r}_{i}-\textbf{r}_{j}|$.
In our numerical implementation, we adopt the Plummer
softening kernel \citep[e.g.][]{Galacticdynamics} with
softening length $1.7\times 10^6~{\rm cm}$, which is
equivalent to the Schwarzschild radius for a $5.75~\Msol$ BH.

We add post-Newtonian terms ${\bf{a}}_{\rm PN}$ to
Eq.~(\ref{eq:Newton-pairs}) up to order 2.5, which includes the loss
of orbital energy and angular momentum via gravitational waves (GWs).
The full expressions for these terms can be found in, e.g.,
\citet[][see also \citealt{damour1981}]{Kupi+06}.  At sufficiently
small pair separations, the orbital decay due to GW emission leads to
merger on a timescale $\propto a^4$ where $a$ is the semimajor axis \citep{Peters63, Peters64}.  It
turns out that no BHs in our simulations reach separations where GW
emission is relevant for their orbital evolution\footnote{
For reference, GW emission becomes the dominant orbital
evolution mechanism at an orbital distance $a< 10^{-5}\pc$
for a circular binary consisting of two BHs with masses $10^{5} \Msol$ and $100 \Msol$.
We find that the central pair is rarely disrupted once
they reach a separation $\sim 1\pc$, as described later in the text (\S\ref{subsec:finalmasses}).}
Notice that This finding is
different from the similar, earlier studies by \cite{Tagawa+15,
  Tagawa+16}, who considered a cluster of BHs with a smaller initial
separation, and found that the post-Newtonian terms can sometimes be
important and lead to BH-BH mergers.  We defer the discussion of this
point to \S~\ref{sec:discuss} below.\\

\item \noindent\textit {Dynamical friction and gas drag}\\
An object in motion through a medium creates an overdensity, or wake,
behind it, whose gravitational pull acts as a dissipative drag on the
object's motion.  In this study, we consider dynamical friction due to
both DM and gas.

For the DM contribution, we adopt the standard Chandrasekhar
formula\citep{Galacticdynamics},
\begin{equation}\label{eq:df1}
\textbf{a}_{{\rm df},i}^{\rm (DM)}=-4\pi ~\ln\Lambda ~f(X_{i})
~ \frac{G^{2} M_{i}}{v_{i}^{3}}~\rho_{\rm
  DM}(r_{i})~\textbf{v}_{i},
\end{equation}
with
\begin{equation}\label{eq:df2}
f(X_{i})\equiv {\rm erf}(X_{i})-\frac{2}{\sqrt{\pi}} ~X_{i}
~\exp\left(-X_{i}^{2}\right),
\end{equation}
where $X_{i}\equiv v_{i}/(\sqrt{2} \sigma_{v})$ and $\sigma$
is the velocity dispersion, $\simeq c_{\rm s}$. We use
$\ln\Lambda=3.1$ and indicate with $\rho_{\rm DM}(r_{i})$ the
DM density at the location of the $i$-th BH.
		
For the gas, we adopt the following formula from \cite{dyformula1},
which incorporates behaviours found in numerical simulations for
subsonic and supersonic regimes \citep{dyformula2, dyformula3,Chapon+13}.  In
our implementation, the specific drag force vector always points
opposite to the direction of BH motion, and is given by:
\begin{equation}\label{eq24}
\textbf{a}_{{\rm df},i}^{\rm (gas)}=-4\pi ~G^{2}~M_{i}~\rho_{\rm
  gas}(r_{i}) ~\frac{1}{v_i^{3}}\times f^{\rm (gas)}({\mathcal
  M}_i)\textbf{v}_{i},
\end{equation}
with 
\begin{equation}\label{eq25}
f^{\rm (gas)}({\mathcal M}_i)=\begin{cases} 0.5~\ln\Lambda
\Big[{\rm erf}\left (\frac{{\mathcal M}_i}{\sqrt{2}}\right)
  -\sqrt{\frac{2}{\pi}}{\mathcal M}_i~
  \exp\left(-\frac{{\mathcal M}_i^{2}}{2}\right)\Big]
\\ \hspace{100pt} 0\leqslant {\mathcal M}_i \leqslant
0.8;\\ 1.5~\ln\Lambda \Big[{\rm erf}\left (\frac{{\mathcal
      M}_i}{\sqrt{2}}\right) -\sqrt{\frac{2}{\pi}}{\mathcal
    M}_i~ \exp\left(-\frac{{\mathcal M}_i^{2}}{2}\right)\Big]
\\ \hspace{93pt} 0.8\leqslant {\mathcal M} \leqslant
\mathcal{M_{\rm eq
}};\\ \frac{1}{2}\ln\Big(1-\frac{1}{{\mathcal
    M}_i^{2}}\Big)+\ln\Lambda \\ \hspace{115pt} {\mathcal M}_i
> \mathcal{M_{\rm eq}}.\\
\end{cases}
\end{equation}
Above, ${\mathcal M}_i\equiv v_{i}/c_{\rm s}$ is the Mach
number of the $i$th BH, and $c_{\rm s}$ is the isothermal
sound speed of the gas.  We use $\ln\Lambda=3.1$, the same as
for the DM.  The corresponding value of $\mathcal{M_{{\rm
      eq}}}$ that makes the above function continuous with
respect to ${\mathcal M}$ is approximately $1.8$.

With the density distributions we use in our model (see 
\S~\ref{subsec:model} above), and in particular near the center of the
model protogalaxy, the effects of gas dominate over that of DM---both
in dynamical friction and background gravitational force.  Although we
include the DM-related force calculations for completeness, they do
not play a major dynamical role.

The expressions for dynamical friction given above were derived under
the assumption of non-accelerated motion in a uniform density
distribution.  Capturing the effects of nonlinear dynamical friction
along an accelerated trajectory, in a non-uniform background medium,
would require hydrodynamical simulations, including the self-gravity
of the surrounding medium. This is outside the scope of this paper,
but we note that existing studies of dynamical friction in a
nonuniform medium or for perturbers on nonlinear trajectories
\citep[e.g][]{Brandenburg01,Just05,Kim07,Kim10}, do not report major
differences from the Chandrasekhar formula.  We therefore simply
evaluate the formulae given above, by using the value of the density
at the coordinates of each BH; this is the typical approach taken in
similar numerical studies \citep[e.g.][]{Blecha+11,Guedes+11}.  For
comparison, we have run a second set of simulations in which the
dynamical friction forces were computed by averaging the density
values at a distance around each BH; we defer discussing the details
of this comparison until \S\ref{subsec:modelsummary} below. \\
	
\item\noindent\textit{Gravity of the background matter}\\
The background gas and DM exert a gravitational pull on the BHs.
Because we assume a spherically symmetric density profile, this force
points toward the center of the potential.  It can be expressed as
\begin{equation}
\label{eq:bg}
\textbf{a}_{{\rm bg}, i}=-\frac{G~M_{{\rm bg},i}}{r_{i}^{3}} \textbf{r}_{i},
\end{equation}
where $\textbf{r}_{i}$ is vector pointing from the 
center of the halo to the $i$-th BH and $M_{{\rm bg},i}$ is
the mass of ambient gas and DM contained inside $r<r_i$.

Our assumption that the background matter distribution remains static
will fail when $M \gtrsim M_{\rm bg}$, i.e. when the BH mass exceeds
that of the matter inside its orbit. In this case, the matter at the
center will be strongly perturbed by the BH, and our prescription of a
static background is invalidated.  Treating the dynamical reaction of
the gas and DM distribution to a massive BH is beyond the scope of our
computational methods. However, in order to assess the possible impact
of this assumption on our results, we have run two sets of simulations
with different treatments of the background force.  In the first set,
we treat this force simply as given by equation \ref{eq:bg}.  In the
second, we set it to zero if $M > M_{\rm bg}$.  The full set of our
simulations is described in \S\ref{subsec:modelsummary}.\\

\item \noindent\textit{Accretion-induced deceleration}\\ 
The BH decelerates 
through conservation of linear momentum,  
\beq
\textbf{a}_{{\rm acc},i}=-\frac{\dot{M}_i}{M_i}\textbf{v}_i.
\label{eq:a_acc}
\eeq
	
\end{enumerate}

\subsection{Accretion rate}
\label{subsec:mdot}

We next detail our prescription for the mass growth of each BH due to
gas accretion.  We base our model on the recent numerical study by
\citetalias{Inayoshi+16}, whose key finding is that the accretion rate
can significantly exceed the Eddington rate (see also the other
references mentioned in the Introduction). \citetalias{Inayoshi+16}
found that spherically symmetric BH accretion solutions with radiative
feedback were divided into several qualitatively distinct regimes,
depending on the ratio of the Bondi accretion rate of ambient gas ($\rho_{\infty}$),
\begin{equation}\label{eq:Bondi}
\dot{M}_{\rm B}=\frac{4\pi G^{2}M^{2}~\rho_{\infty}}{c_{\rm s, \infty}^{3}},
\end{equation}
 ($c_{\rm s,\infty}$ being the sound of speed of the ambient gas), 
to the Eddington rate\footnote{Our notation was chosen to match that
  of \citetalias{Inayoshi+16}. Many other works define $\dot{M}_{\rm
    Edd}$ to correspond to the Eddington luminosity with a radiative
  efficiency of $\eta$, which would be $\dot{M}_{\rm Edd}/\eta$, i.e.
  a factor of 10 higher than eq.~(\ref{eq:Edd}) for $\eta=0.1$.}
\begin{equation}\label{eq:Edd}
\dot{M}_{\rm Edd}\equiv \frac{L_{\rm Edd}}{c^2}=\frac{4\pi G M}{\kappa_{\rm es}~c}=2.2\times10^{-9}\frac{M}{\Msol}\Msol~\yr^{-1},
\end{equation}
where $L_{\rm Edd}$ is the Eddington luminosity and $\kappa_{\rm es}$ is the Thomson scattering opacity.

\citetalias{Inayoshi+16} concluded that:\\
(i) Under conditions where the ratio of the canonical Bondi rate to
the Eddington rate $\dot{M}_{\rm B}/\dot{M}_{\rm Edd}\la 0.1-1$, the
BH accretion rate is $\dot{M}\sim \dot{M}_{\rm B}$;\\
(ii) If $1\la \dot{M}_{\rm B}/\dot{M}_{\rm Edd}\la 100$,
photoionization by the light produced by the accretion flow causes the
accretion onto the BH to be intermittent, with a time-averaged rate
$\dot{M}\la \dot{M}_{\rm Edd}$; \\
(iii) If $\dot{M}_{\rm B}/\dot{M}_{\rm Edd}\ga 3000$, the large ram
pressure of the inflowing gas, combined with photon trapping below the
photosphere, renders radiative feedback ineffective, and accretion
proceeds unimpeded at $\dot{M}\sim \dot{M}_{\rm B}$
\citep[cf.][]{Begelman12}.\\
In the intermediate regime between cases (ii) and (iii) above, i.e. if
$100\la\dot{M}_{\rm B}/\dot{M}_{\rm Edd}\la 3000$, the accretion rate
remains uncertain, because of the unresolved role of hydrodynamical
instabilities.

Our accretion prescription closely follows the behaviour outlined
above, with a few modifications.  First, we replace the Bondi rate
for a stationary mass with the Bondi-Hoyle-Lyttleton (BHL) rate, to account
for the fact that the BHs in our simulation are in motion with respect
to the surrounding gas.  Second, whereas the canonical expression for
the Bondi accretion rate (eq. \ref{eq:Bondi}) uses the ambient density
``at infinity'' $\rho_\infty$, we instead use the value of the density
averaged over the spherical region around the BH defined by the Bondi
radius
\begin{equation}\label{eq:rB}
r_{{\rm B},i}=\frac{2GM_{i}}{c_{\rm s}^{2}+v_i^{2}}.
\end{equation}
The resulting expression for our modified Bondi rate is
\begin{equation}\label{eq:BHL}
\dot{M}_{{\rm B},i}=\frac{4\pi G^{2}M_{i}^{2}~\langle\rho_{\rm B}(r_i)\rangle}{c_{\rm s}^{3}(1+\mathcal{M}_i^{2})^{3/2}},
\end{equation}
where $\langle\rho_{\rm B}(r_i )\rangle$ denotes the aforementioned
average density of gas at the surface of the ``Bondi sphere'' around the BH.

Third, we conservatively assume that the BH accretion rate should not
be higher than the mass inflow rate into the center of the halo from
larger scales, as this would deplete the central gas density, without
the possibility of a steady replenishment from larger radii.  In pristine
atomic--cooling haloes, the hydrodynamical simulations mentioned above
typically find this inflow rate to be
\beq
\dot{M}_{\rm in} = \frac{c_{\rm s}^3}{G} \approx 0.5 \Msol \yr^{-1}.
\label{eq:Minflow}
\eeq
A possible caveat here is that the inflow rate in the presence of
metal and/or ${\rm H_2}$ cooling may be $\approx$ two orders of
magnitude lower. On the other hand, the rate increases steadily as the
halo grows in mass, and recent simulations have found that pressure
and gravitational torques can maintain $\sim \Msol \yr^{-1}$ inflow
rates down to $\sim$pc scales, even in the face of radiative cooling
and SN feedback \citep{PrietoEscala2015}.

Our implementation of the \citetalias{Inayoshi+16} accretion regimes
can therefore be summarized as:
\beq
\label{eq:acc}
\dot{M}=
\begin{cases}
\min[\dot{M}_{\rm B},\frac{1}{\eta}\dot{M}_{\rm Edd}, \dot{M}_{\rm in}]\\
\qquad\qquad({\rm if}~{\rm min}[\dot{M}_{\rm B},\dot{M}_{\rm in}]<3000 \dot{M}_{\rm Edd})\\
$~$\\
\min[\dot{M}_{\rm B}, \dot{M}_{\rm in}]\\
\qquad\qquad({\rm if}~{\rm min}[\dot{M}_{\rm B},\dot{M}_{\rm in}]\ge 3000 \dot{M}_{\rm Edd})
\end{cases}.
\eeq

\begin{table*}
\centering
\caption{Summary of our 6 prescriptions for BH accretion and 2
  different treatments of the central background potential adopted in
  our simulations. These constitute a set of 12 models.  Note that the
  accretion prescription in Eq.~\ref{eq:acc} follows
  \citetalias{Inayoshi+16}. In model ``I'' this prescription is
  adopted independently of whether the BH mass exceeds or not the mass
  contained within its Bondi radius, $M_{\rm B}$, whereas in the
  ``$f_{\rm in}$'' models the accretion rate is capped to a fraction
  $f_{\rm in}$ of the inflow rate in the self-gravitating regime, when
  $ M_{\rm B}\ge M $. Model ``E'' denotes a commonly adopted
  Eddington-limited accretion prescription, and in the $\dot{M}=0$
  reference case we do not allow any accretion.}
\setlength\extrarowheight{5pt}

\begin{tabulary}{1\linewidth}{| l | c | c | c | }
\hline
\multicolumn{2}{|c|}{Model} &	$\dot{M}$ when $M_{\rm B}<M$ & 	$\dot{M}$ when $M_{\rm B}\geq M$ \\ %\cline{3-4}	
				\hline
	&$f_{\rm in}=1$ & eq.\ref{eq:acc} & $\dot{M}=f_{\rm in}\dot{M}_{\rm in}=\dot{M}_{\rm in}$ \\
	     &$f_{\rm in}=10^{-3}$ & eq.\ref{eq:acc} & $\dot{M}=10^{-3}\times\dot{M}_{\rm in}$ \\
	     	 	Accretion &$f_{\rm in}=0$ & eq.\ref{eq:acc} & $\dot{M}=0$ \\
	                                    Prescription      & I & eq.\ref{eq:acc} &  eq.\ref{eq:acc} \\
	                                       & {E} & $\min[\dot{M}_{\rm B},\frac{1}{\eta}\dot{M}_{\rm Edd}, \dot{M}_{\rm in}]$ &   $\min[\dot{M}_{\rm B},\frac{1}{\eta}\dot{M}_{\rm Edd}, \dot{M}_{\rm in}]$\\
	                                      & $\dot{M}=0$ & $\dot{M}=0$ &  $\dot{M}=0$ \\
	         \hline
	         \hline
		& & \multicolumn{2}{c|}{}  \\
	     	  Background   &  On &  \multicolumn{2}{c|}{$a_{{\rm bg},i}$ always on} \\
	               Potential   &  Off &  \multicolumn{2}{c|}{$a_{{\rm bg},i}=0$ if $M_i>M_{\rm bg}(<r_i)$} \\
		& & \multicolumn{2}{c|}{}  \\
	                                      	                                      \hline
	\end{tabulary}
	\label{table:modelsummary}
\end{table*}

In this study, we take $\eta=0.1$. We note that this may
somewhat overestimate the accretion rate in the Eddington-limited
regime, as both \citet{ParkRicotti12} and \citetalias{Inayoshi+16}
found that the time-averaged rate is limited to $\sim0.5 \dot{M}_{\rm
  Edd}$.  We implement Equation (\ref{eq:acc}) as long as the BH
dominates the gravitational potential inside its Bondi sphere,
i.e. $M>M_{\rm B}$.  On the other hand, if $M<M_{\rm B}$, the Bondi
formalism breaks down.  The latter condition roughly coincides with
the gas inside the Bondi sphere becoming self-gravitating and
Jeans-unstable.  In this regime, the accretion rate is plausibly of
order $\dot{M}_{\rm in}\approx c_{\rm s}^3/G$.  However, how much of
this canonical inflow rate ends up accreting onto the BH remains
uncertain, and will depend on factors such as gas cooling, turbulence
and angular momentum transport, and the BH's specific accelerated
trajectory.  We here parameterize the accretion rate in this
self-gravitating gas regime as $\dot{M}=f_{\rm in}\dot{M}_{\rm in}$,
and consider the two extreme values of $f_{\rm in}=0$ and $1$, as well
as an intermediate value of $f_{\rm in}=10^{-3}$.

To further explore the dependence of our results on this accretion
prescription, we have also run a set of simulations where $\dot{M}$
continues to be given by the Bondi rate even when $M_{\rm B}>M$, and
another set where $\dot{M}$ continues to follow the
\citetalias{Inayoshi+16} prescription described above, regardless of
whether $M_{\rm B}$ is larger or smaller than $M$.

\subsection{Code Description}
\label{subsec:code}

We perform 3-dimensional $N$-body simulations with a 4th-order,
5-stage Runge-Kutta-Fehlberg method (RKF45 method, \citealt{Erwin}),
using adaptive time steps. RKF45 is a highly accurate and stable
integration method among the large class of Runge-Kutta schemes,
particularly by adapting the Butcher tableau for Fehlberg's 4(5)
method.

We solve equation (\ref{eq:eom}) as described in the preceding text,
and at each time step we update the components of the position,
velocity and acceleration, as well as the masses of the BHs, according
to prescribed forces and accretion rates.  To ensure numerical
precision, our computational scheme varies the value of each
subsequent time step analytically, so that numerical errors for each
variable in the simulation do not exceed $10^{-13}$ times the size of
the variable.  In some cases, however, this algorithm can spend an
excessive amount of time calculating trivial interactions. In order to
avoid such situations and to achieve acceptable code speeds, we
implement a shortcut in the form of a minimum time step, \beq \Delta
t_{\rm short} = 10^{-6} \times {\rm min}\{\tau_{{\rm dyn},ij},
\tau_{{\rm df},i}, M_{i}/\dot{M}_i, v_i/a_i \}, \eeq where
$\{\tau_{{\rm dyn},ij} \}$ is the set of dynamical times evaluated for
each BH pair, as well as for each BH and the background potential;
$\tau_{{\rm df},ij}$ is the timescale for orbital energy dissipation
by dynamical friction and gas drag for each BH pair; $M_{i}/\dot{M}_i$
is the accretion timescale, and $v_i/a_i$ is the ratio of the speed to
the net acceleration for each BH.

The main body of the code is the same as the one that was used
in \cite{Ryu+16}. We refer the reader to that paper for further details.

\begin{figure*}
\centering 
{\includegraphics[width=8.0cm]{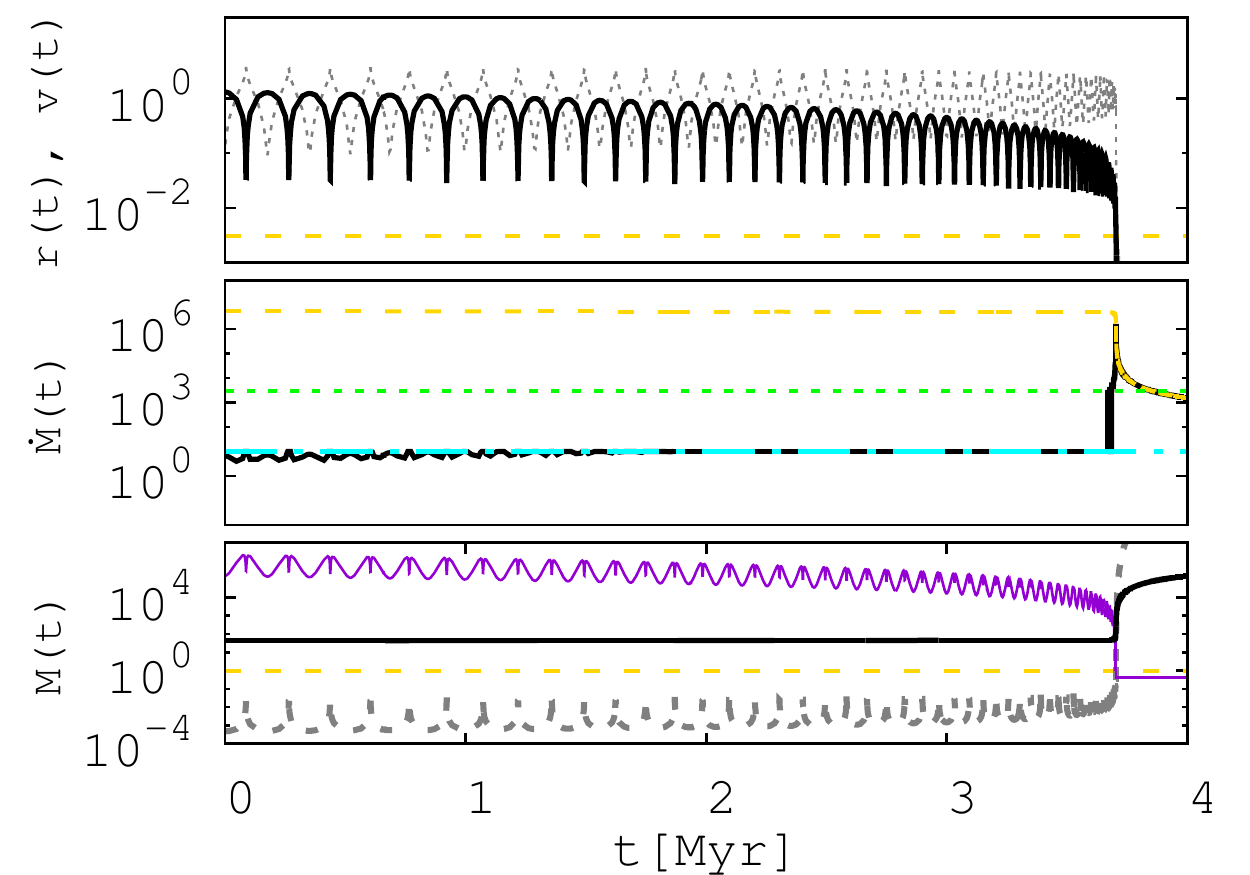}}
{\includegraphics[height=5.5cm,width=1.5cm]{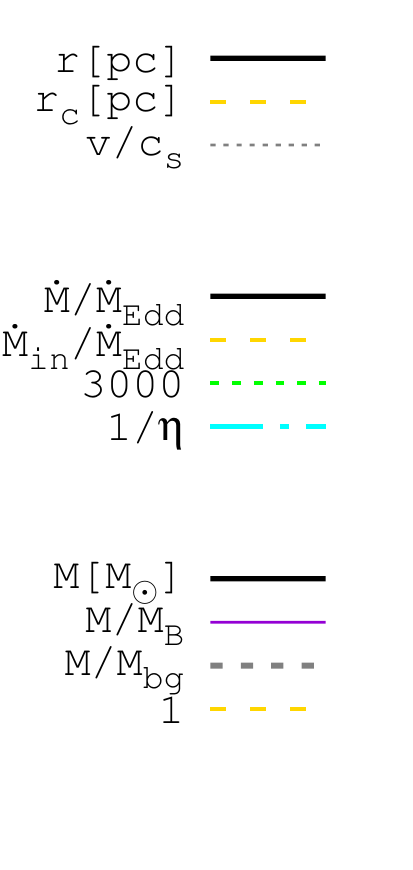}}
{\includegraphics[width=8.0cm]{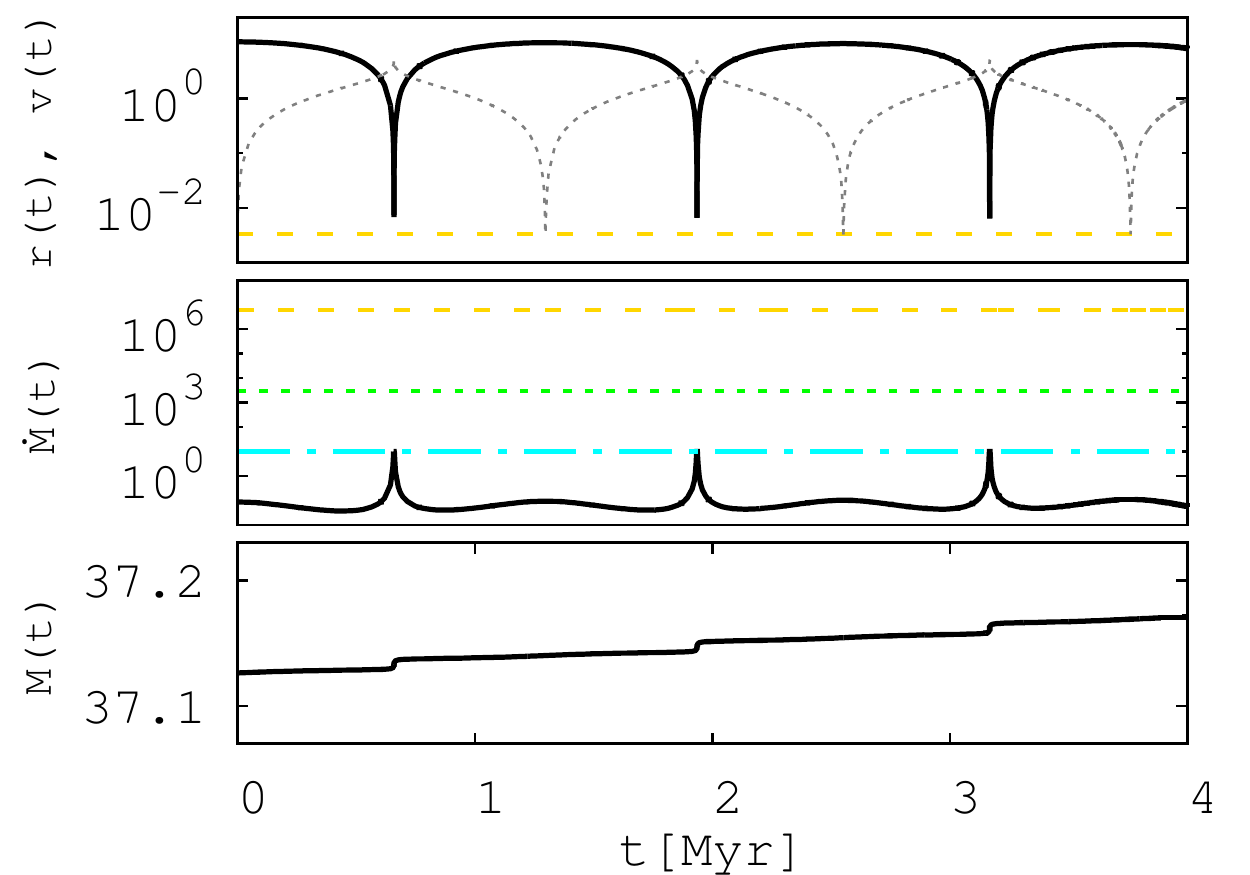}}
\caption{\textit {Top left panel:} The radial distance $r$ (black
  solid line), and Mach number (gray dotted line) of a BH which
  experiences the hyper-Eddington accretion as it sinks to the halo
  core, whose size $r_{\rm c}$ is also displayed for reference (yellow
  dashed line) in Model "$f_{\rm in}=1$'' with $a_{\rm bg}$ 'On' (see
  Table~\ref{table:modelsummary}).
  \textit {Middle left panel:} The corresponding accretion rate
  $\dot{M}$ of the BH in Eddington units (solid black line). For
  reference, the panel also shows the the self-gravitating inflow rate
  $\dot{M}_{\rm in}$ (yellow dashed) also in Eddington units, as well
  as the critical values at which the accretion switches modes
  [i.e. to Eddington (blue dot-dashed) or hyper-Eddington (green
    dotted)].
 \textit{Bottom left panel:} The corresponding mass of the BH is shown
 (solid black line), together with the mass $M_{\rm B}$ enclosed
 within its Bondi radius (solid magenta) and the gas mass $M_{\rm bg}$
 contained inside its orbit (dashed gray). The accretion rate and the
 BH mass both grow rapidly once the BH sinks the core.  At the same
 time, the BH begins to dominate the central potential, but not the
 mass inside its Bondi sphere.
 \textit {Right panels:} The above behaviour is contrasted with a BH
 on a highly elliptical orbit but with a larger semimajor axis, which
 never makes it inside the core. This BH never experiences
 (hyper-)Eddington accretion, and its mass remains near its initial
 value.  }
\label{fig:3}
\end{figure*}

\subsection{Summary of simulation sets}
\label{subsec:modelsummary}

Our fiducial set of models implement the prescriptions described
above.  The accretion rate onto the BH is determined by the
\citetalias{Inayoshi+16} accretion rates (eq. \ref{eq:acc}) if
$M>M_{\rm B}$, and $\dot{M}=f_{\rm in}M_{\rm in}$ in the
self-gravitating regime, once $M\le M_{\rm B}$.  As noted above, we
have run two additional sets of simulations, in which $\dot{M}$ is
either given by Eq.~(\ref{eq:acc}) regardless of how $M$ compares to
$M_{\rm B}$ (which we will refer to as (Prescription ``I''; for
Inayoshi et al.)  and another in which BH accretion tracks the Bondi
rate, but is always Eddington-limited, i.e. $\dot{M}=\min[\dot{M}_{\rm
    B},\frac{1}{\eta}\dot{M}_{\rm Edd}, \dot{M}_{\rm in}]$
(Prescription ``E'', for Eddington).  Note that the latter is a
prescription commonly adopted in numerical and semi-analytic studies
of BH growth in the early Universe (see reviews by,
e.g. \citealt{Volonteri10,Haiman13} and references therein) As a
simple control, we have also run simulations with no accretion.

For each of the six accretion prescriptions listed above, we consider
a case where the background gravitational force $a_{{\rm bg},i}$ is
always present and points inward.  We then consider a
second case, motivated in the previous subsection, where $a_{{\rm
    bg},i}$ is set to zero whenever the BH is more massive than the
mass enclosed inside its present position [i.e. if $M_i>M_{\rm
    bg}(r_i)$].

Our full suite of simulations is summarized in Table
\ref{table:modelsummary}. Each of the twelve models we have described
above (six different accretion prescriptions, and two different
treatments of the background gravitational force at small radii) were
simulated multiple times using different initial values for BH masses,
positions, and velocities (the criteria for our initial conditions are
described in \S\ref{subsubsec:BHsetup} and \ref{tab1}).  We simulated
each model using 43 distinct sets of initial conditions; each set of
initial conditions was recycled 12 times, using the different
prescriptions in the 12 different models.

\section{Results}

We now turn to the results of our $N$-body simulations.  We briefly
summarize our major findings below, and follow these with detailed
explanations and analyses.

\begin{enumerate}
\item We found that in a majority (24 out of 43) of initial
  condition sets, an IMBH of mass $\sim 10^3$ to $\sim 10^5$ $\Msol$
  formed as a result of hyper-Eddington accretion, in all the models
  that allowed for this accretion mode (i.e. in models ``$f_{\rm
    in}=1$'',``$f_{\rm in}=10^{-3}$'',``$f_{\rm in}=0$'' and ``I''
  listed in Table 2).
\item If one set of initial conditions results in IMBH formation in
  one hyper-Eddington model, then it does so in all the others.
  The determining factor is whether the BH passes through
    a dense, gas-rich region (as a result of small semimajor axis,
    small pericenter, or both) where dissipation of orbital energy via
    gas drag is efficient.
\item All of the IMBHs end up within the central $\la$ 0.01~pc
  of the protogalactic halo, strongly suggesting that they are viable
  precursors of nuclear SMBHs observed as quasars at $z>6$.
\item There is at most one IMBH in each simulation; we do not find a
  single instance of multiple IMBHs forming.
\item Many of the IMBHs capture a lighter BH into a close, sub-parsec
  orbit, and we argue that such systems could lead to EMRI events
  detectable by planned gravitational-wave observatories, such as {\it
    eLISA}.
\item We do not find mergers between stellar-mass BHs, in contrast to
  similar studies by \cite{Tagawa+15,Tagawa+16}.  The main reason for
  this appears to be simply the larger radii at which we initially
  place the BHs.
\item The above findings appear to be robust with respect to our
  treatment of both dynamical friction and the gravitational force due
  to the background matter distribution.
\end{enumerate}

\subsection{The onset of hyper-Eddington accretion}
\label{StoIMBH}

The first significant event in our simulations is the descent of the
innermost BH to the dense gaseous core, which is driven by the decay
of its orbit due to dynamical friction.  Because the BHs are
relatively widely separated in our initial conditions (the initial mean separations are
$\simeq 75\pc$), three-body
interactions at this stage are rare.  As the innermost BH sinks even
closer to the center, its ambient density increases and its Bondi
accretion rate increases.  The accretion rate eventually transitions
from $\dot{M}=\dot{M}_{\rm B}<\dot{M}_{\rm Edd}/\eta$ (sub-Eddington
Bondi), to $\dot{M}=\dot{M}_{\rm Edd}/\eta < \dot{M}_{\rm B}$
(Eddington-limited), and finally (in 25 out of 43 cases) to
$\dot{M}=\dot{M}_{\rm B} > 3000~\dot{M}_{\rm Edd}$ (hyper-Eddington;
accretion unimpeded by radiation feedback), as dictated by
Eq.~(\ref{eq:acc}).  This qualitative picture is shared by all of our
simulations in which hyper-Eddington accretion occurs.

This progression is illustrated in the left panel of Figure~\ref{fig:3},
which shows the journey in position and mass for a BH that undergoes
hyper-Eddington accretion.  The data is taken from a simulation run
using the $f_{\rm in}=1$ accretion scenario and with the background
gravitational force always present---however, we reiterate that the
behaviour shown here is shared by all examples of hyper-Eddington
accretion in our simulations.

The top left panel of this figure shows the position of the innermost
BH with the core radius $r_{\rm c}$ shown for reference.
In the middle panel, we show the BH accretion rate in units of
$\dot{M}_{\rm Edd}$, alongside the two critical values that determine
the accretion regime according to Eq.~(\ref{eq:acc}):
$3000~\dot{M}_{\rm Edd}$, and $\dot{M}_{\rm in}$.
Finally, in the bottom panel, we show the mass $M$ of the BH (in
$\Msol$).  We also plot $M/M_{\rm B}$, the ratio of the BH mass to the
gas mass inside its Bondi sphere.  Recall that when this ratio is
greater than unity, we study evolution under different accretion
prescriptions (the difference between our $f_{\rm in}=1$, $f_{\rm
  in}=10^{-3}$, $f_{\rm in}=0$, and ``I'' models; see
Table~\ref{table:summary2}).  Further, we show $M/M_{\rm bg}$, the
ratio of the BH mass to the mass contained in the halo inward of the
BH's radial position.  If this ratio is greater than unity, we also
study BH evolution without the background gravitational in the
``Background Potential Off'' simulations (see
Table~\ref{table:summary2}).

For comparison, on the right side of the figure, we have plotted the
same information for another BH in the same simulation that does
\textit{not} undergo hyper-Eddington accretion.  This BH is on a
highly elliptical orbit around the center of the halo, but its Bondi
accretion value never exceeds the $\dot{M}_{\rm B} = 3000~\dot{M}_{\rm
  Edd}$ threshold to overcome radiative feedback.  

In the left side of this figure, we see that it takes the accreting BH
only a few Myr to sink to the center where it begins to undergo
hyper-Eddington accretion.  For BHs that undergo hyper-Eddington
accretion in our simulations that allows this (models $f_{\rm in}=1$,
$f_{\rm in}=10^{-3}$, $f_{\rm in}=0$, and ``I''), an average of $5.6
\Myr$ elapses between when the Bondi accretion rate first reaches the
Eddington limit and when it reaches $3000\dot{M}_{\rm Edd}$.  This is
much shorter than the Salpeter time ($\sim45\Myr$ for the adopted
radiative efficiency $\eta=0.1$), indicating that the increase in the
Bondi accretion rate is caused by the increase in ambient density
($\dot{M}_{\rm B}\propto \rho_{\rm gas}\propto r^{-2}$ for $r\ga
r_{\rm c}$), as opposed to mass growth ($\dot{M}_{\rm B}\propto M^2$).

\begin{figure}
	%\centering
	\hspace{-0.5in}{\includegraphics[width=11cm]{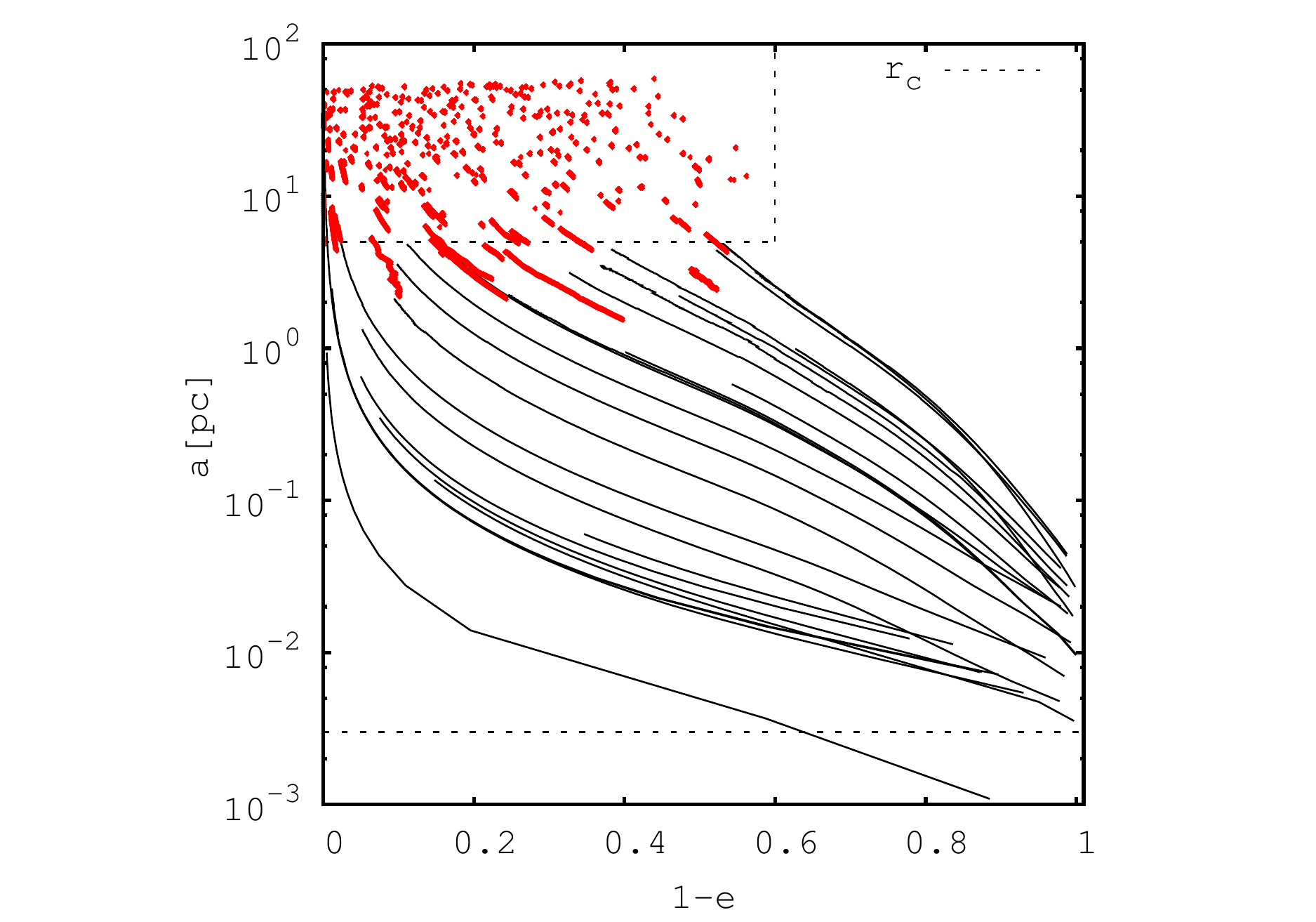}}
	\caption{The evolution of the semimajor axis and eccentricity for 430
		BHs in 43 simulations with the prescription ``$f_{\rm in}=1$,
		$a_{\rm bg}$ on.''  The BHs begin with large semimajor axes and
		high-eccentricity orbits (upper left of the panel).  The red dots
		represent the orbits of 406 BHs that do not sink to the center
		within the simulation runtime of 500 Myr. The black lines represent
		the orbits of the 24 BHs that sink to the dense central region and
		grow into IMBHs.  The orbits circularize as they decay, before
		finally plunging radially to the center (this last phase is not
		shown in the figure). Only BHs with an initial distance of $\la 5$pc
		from the center are found in this category. 
		The dotted box region on the upper left demarcates the region in the
		$a$-$e$ parameter space where we never found examples
		of BHs that successfully grow into a central massive BH.
		The dashed horizontal line marks the core radius at $r_{\rm
			c}=0.003\pc$.  }
	\label{fig:a_vs_e}
\end{figure}

In Figure~\ref{fig:a_vs_e}, we show the semimajor axis (vertical
axis) and eccentricity (plotted as $1-e$, horizontal axis) of each of
our BHs in the ``$f_{\rm in}=1$, $a_{\rm bg}$ on'' simulation set (a
total of 43 runs and 430 BHs, 24 of which grow to become IMBHs).
Because the BH orbits are not Keplerian (and therefore not
elliptical), the concept of eccentricity is not rigorously defined.
We evaluate the instantaneous eccentricity using the standard formula
for Keplerian orbits
\beq 
e=\sqrt{1+\frac{2\epsilon \ell^2}{\{G[M+M_{\rm bg}(r)]\}^2 }},
\eeq
where $\epsilon$ is the specific energy of the orbit (orbital energy
divided by the ``instantaneous reduced mass'' $M~M_{\rm
  bg}(r)/[M+M_{\rm bg}(r)]$), and $\ell$ is the specific angular
momentum (orbital angular momentum divided by the instantaneous
reduced mass).  Note that the gravitational potential used in
calculating $\epsilon$ is logarithmic, and the enclosed central mass
$M_{\rm bg}(r)$ varies with the orbital radius.

The red dots in Figure~\ref{fig:a_vs_e} represent the orbital
evolution of BHs that do not make it to the dense central region to
become IMBHs within the 500 Myr runtime of the simulations; the black
curves represent the orbital evolution of BHs that do grow into IMBHs.
The dotted box in the upper left approximately encloses
  initial orbital parameters ($a$ and $e$) for BHs which do not grow to
  IMBH/SMBH. BHs with the initial $a$ and $e$ outside the box have
  sunk to the core and have experienced hyper-Eddington accretion,
  whereas we find no such examples with initial $a$ and $e$ that lie
  inside the box.  Notice that some BHs with small initial semimajor
  axes (i.e. outside the box) migrate to the center after the first BH
  becomes massive, and form a bound pair with it.  The dashed
  horizontal line near the bottom marks the core radius, $r_{\rm
    c}=0.003\pc$.

BHs evolve from having large semimajor axes and high-eccentricity
orbits (upper left of the panel) about the center of the protogalaxy
potential, to having tighter, nearly circular orbits (lower right
portion of the panel).  As a BH approaches the center of the
protogalaxy, it sinks to the center more quickly than it can complete
an orbit.  The final plunge into the center of the protogalaxy is not
plotted, as we find that $e$ cannot be reliably calculated from the
shape of the orbit.
\\

The orbital evolution and the final transition to hyper-Eddington
accretion described above can be understood as follows.  Outside the
core, the mass enclosed inside the BH orbit is $M_{\rm bg}\approx 4\pi
\rho_{\rm c}r_{\rm c}^2 r \propto r$, and the dynamical time can be
expressed as
\beq \tau_{\rm
  dyn}=\left(\frac{r}{a_{\rm g}}\right)^{1/2} \approx 170
\left(\frac{r}{r_{\rm c}}\right) \yr\,.
\label{eq:tdyn}
\eeq

Barring an encounter with another BH, BHs have velocities $v\sim
c_{\rm s}$ or smaller. In this subsonic regime, $f^{\rm gas}v^{-3}\la
c_{\rm s}^{-3}$ in Eq.~\ref{eq25}.  Writing $a_{\rm df} \sim
v/\tau_{\rm df}$, we can estimate $\tau_{\rm df}$ as
\beq
\tau_{\rm df} \sim (1-4)\times10^3 \left(\frac{M}{10\Msol}\right)^{-1} {\rm max}\left[1, \left(\frac{r}{r_{\rm c}}
\right)^{2}\right] \yr\,,
\label{eq:tdf}
\eeq
where the leading factor depends on whether the Mach number is less
than or greater than $\mathcal{M}=0.8$.  The azimuthal force due to
dynamical friction results in circularization and orbital decay on a
timescale of $\tau_{\rm df}$, whereas in the radial direction the
background force dominates over dynamical friction ($\sim \Myr$ for
pericenter passage of $\ga 30~r_{\rm c}$ and $M\sim 30 \Msol$).
Because $\tau_{\rm df}\propto r^2$ outside the core, the decay
accelerates, with the final stages occurring over a few thousand
years.

Since the orbital decay occurs on a timescale much shorter than the
Salpeter time, the BH does not grow significantly by either
sub-Eddington or Eddington-limited accretion.  If $M\la 60\Msol$, then
$r_{\rm B}\la r_{\rm c}$ and the orbital velocity is comparable to or
less than the sound speed.  The Bondi accretion rate for stellar-mass
BHs outside the core can be estimated as
\begin{align}
\dot{M}_{\rm B}&\approx \frac{4\pi G^2 M^2 \rho(r)}{c_{\rm s}^3}\nonumber\\
&=7\times 10^{-3}\left(\frac{M}{10\Msol}\right)^2 \left(\frac{r}{r_{\rm c}}\right)^{-2}\Msol \yr^{-1}
\nonumber\\
&=3.1\times 10^{5} \left(\frac{M}{10\Msol}\right) \left(\frac{r}{r_{\rm c}}\right)^{-2}\dot{M}_{\rm Edd}.
\label{eq:mdotbondi}
\end{align}
Hence an infalling stellar-mass BH begins to undergo hyper-Eddington
accretion when $r\ga r_{\rm c}$.

\subsection{Hyper-Eddington accretion: a brief but dramatic growth spurt}

Once the BH overcomes the $\dot{M}_{\rm B} = 3000~\dot{M}_{\rm Edd}$
threshold, its accretion rate instantaneously increases by a factor of
$3000~\eta=300$.  At the instant after this transition, its mass
growth timescale is $M/\dot{M}\approx 0.1\Myr$ (the Salpeter time
divided by $3000\eta$), but then rapidly shortens as $\propto M^{-1}$
before quickly hitting the ceiling $f_{\rm in}\dot{M}_{\rm in}$
imposed by the large-scale mass inflow rate (eq.~\ref{eq:Minflow}).
For comparison, note that in Eddington-limited growth, $\dot{M}\propto
M$ and the growth timescale is constant at the Salpeter value.  As a
result, when the BH accretion rate becomes hyper-Eddington, its mass
shoots up from a ${\rm few}\times 10\Msol$ to more than $10^3\Msol$ in
less than $\sim 0.1\Myr$.

This rapid mass growth results in two significant transitions in our
simulations.  First, the gas mass inside the Bondi radius exceeds that
of the BH mass.  Again, how the BH is assumed to accrete mass when
$M_{\rm B}>M$ is what distinguishes our $f_{\rm in}=1$, $f_{\rm
  in}=10^{-3}$, $f_{\rm in}=0$ and ``I" models.

If $M\la 60\Msol$, then the Bondi radius is small and the density at
the surface of the Bondi sphere is close to the local density. Then,
considering the Bondi-like profile ($\sim r^{-3/2}$) inside the Bondi sphere,
\begin{align}
M_{\rm B} &\sim \int_0^{r_{\rm B}} \langle\rho_{\rm B}(r')\rangle\left(\frac{r_{\rm B}}{r'}\right)^{3/2}~r'^2~{\rm d}r'\\
&\sim\frac{8\pi}{3}r_{\rm B}^{3}\rho(r)\nonumber\\
&\sim 0.7\Msol \left(\frac{M}{10\Msol}\right)^3 {\rm min}\left[1,\left(\frac{r}{r_{\rm c}}\right)^{-2}\right],
\end{align}
and $M_{\rm B}<M$.

However, since $r_{\rm B}\propto M$, at larger BH masses the Bondi
sphere quickly becomes larger than the gaseous core.  We just
established above that the BH typically begins its hyper-Eddington
accretion near the core.  Therefore, once a BH grows to $M\ga 60
\Msol$, $r_{\rm B}>r_{\rm c}$ and the gas density at the surface of
the Bondi sphere is essentially given by the halo gas profile
evaluated at $r=r_{\rm B}$.  Then we can write
\beq
M_{\rm B} \sim \frac{8\pi}{3}r_{\rm B}^{3}\rho(r_{\rm B})
\approx \frac{8\pi}{3}\rho_{\rm c}r_{\rm c}^2 r_{\rm B}
=2.4 M>M.
\eeq
We conclude that as soon as hyper-Eddington accretion begins, the gas
enclosed inside the Bondi sphere shoots above the BH mass.

The second transition that occurs as the BH grows is that the BH mass
exceeds the mass of the matter enclosed inside its orbit around the
center of the halo ($M_{\rm bg}$).  For $r\gg r_{\rm c}$, the enclosed
mass is simply
\beq
M_{\rm bg}\ga 4\pi \rho_{\rm c}r_{\rm c}^2 r \approx 200 \Msol \left(\frac{r}{r_{\rm c}}\right),
\eeq
whereas inside the core ($r\ll r_{\rm c}$)
\beq
M_{\rm bg}= \frac{4\pi}{3} \rho_{\rm c} r^3 \approx 70 \Msol \left(\frac{r}{r_{\rm c}}\right)^3.
\eeq
Either way, as the BH grows beyond several $100\Msol$ near or inside
the core, our simulations always result in $M>M_{\rm bg}$.  Since the
BH dominates the central potential, the innermost gas distribution
will be strongly disturbed, and will correspond to a radial force
towards the center of the halo.  This is the motivation for running a
second set of simulations, in which the gravitational force of the
background matter is turned off if $M>M_{\rm bg}$.

The above caveats aside, the hyper-Eddington rate initially follows
the Bondi rate and scales as $\dot{M}\propto M^2$.  One significant
aspect of this mode of growth is that, in addition to the raw
accretion rate being much higher than Eddington-limited growth, the
accretion timescale decreases (i.e. the growth rate accelerates and is
faster than exponential).  However, this growth does not last long in
our simulations, because $\dot{M}$ encounters one of two
upper limits.

The first upper limit is $f_{\rm in}\dot{M}_{\rm in}$ (where
$\dot{M}_{\rm in}=c_{\rm s}^{3}/G$), the parameterized gas supply rate
into the center of the halo in the ``$f_{\rm in}=1$'',``$f_{\rm
  in}=10^{-3}$'', and ``$f_{\rm in}=0$'' models.

The second one is due to the fact that as $M$ increases, $r_{\rm B}$
increases, and the density at the surface of the Bondi sphere
decreases.  For $r_{\rm B}\gg r$ and $r_{\rm B}\gg r_{\rm c}$, the
density at the Bondi sphere surface becomes \beq \rho(r_{\rm B})
\approx \rho_{\rm c} \left(\frac{r_{\rm B}}{r_{\rm c}}\right)^{-2}
=\frac{\rho_{\rm c}r_{\rm c}^2 c_{\rm s}^4}{4G^2 M^2}.  \eeq Then the
Bondi accretion rate evaluates to
\begin{equation}
\label{Bondiasymptoticvalue}
\dot{M}_{\rm B}=\frac{4\pi G^{2}M^{2}\rho(r_{\rm B}) }{c_{\rm s}^{3}(1+\mathcal{M}^{2})^{3/2}}
\approx \pi \rho_{\rm c}r_{\rm c}^{2} c_{\rm s}\approxeq0.2\Msol\yr^{-1},
\end{equation}	
or $\approx 40~\%$ of $\dot{M}_{\rm in}$ (=$c_{\rm s}/G^{3}$). 

\begin{table*}
	\centering
	\caption{Average mass, mass ratio, eccentricity ($e$) and
          accretion rate for the first BH-BH pair when its semimajor
          axis is $a\approx1\pc$. Accretion rates are in units of the
          Eddington rate. The subscript ``1'' refers to the more massive
          BH, and ``2'' to the less massive BH. }
        \setlength\extrarowheight{5pt}
	
	\begin{tabulary}{1\linewidth}{c | c |c| c| c| c| c| c }
		\hline
		$a_{\rm bg}$&  Model& $M_{1}[\Msol]$& $M_{2}[\Msol] $& $q$ (=$M_{2}/M_{1}$) &$e$ &$\dot{M}_{1}/\dot{M}_{\rm Edd,1}$ & $\dot{M_{2}}/\dot{M}_{\rm Edd,2}$\\
		\hline
		\hline
		\multirow{6}{*}{on}& $f_{\rm in}=0$ & 160 & 71 & 0.45 & 0.92 & 0 & 38000\\
		& $f_{\rm in}=10^{-3}$ & $1.5\times10^{4}$ & 45 & $3.0\times10^{-3}$ & 0.99& 19 & 4.8\\
		& $f_{\rm in}=1$ & $6.1\times10^{5}$ & 31 & $5.1\times10^{-5}$ & 0.90& 760 & $8.6\times10^{-2}$\\
		& I & $2.5\times10^{5}$ & 38 & $1.5\times10^{-4}$ & 0.87& 10 & 2.9\\
		& E & 450 & 50 & $0.11$ & 0.99& 10 & 5.9\\
		& $\dot{M}=0$ & $41$ & 37 & $0.92$ & 0.96 & 0 & 0\\
		\hline
		\multirow{6}{*}{off}& $f_{\rm in}=0$ & 170 & 91 & 0.55 & 0.92 & 0 & 58000\\
		& $f_{\rm in}=10^{-3}$ & $2.6\times10^{4}$ & 44 & $1.7\times10^{-3}$ & 0.99& 17 & 3.2\\
		& $f_{\rm in}=1$ & $4.1\times10^{6}$ & 29 & $7.0\times10^{-6}$ & 0.84 & 240 & $2.1\times10^{-2}$\\
		& I & $2.7\times10^{6}$ & 33 & $1.2\times10^{-5}$ & 0.98 & 10 & 0.78\\
		& E & $370$ & 53 & $0.14$ & 0.90 & 10 & 6.8\\
		& $\dot{M}=0$ &  41& $37$ & $0.92$ & 0.87 & 0 & 0\\
		\hline
	\end{tabulary}
	\label{table:summary2}
\end{table*}

\vspace{\baselineskip}
To summarize, the dynamics in our simulations evolves according to the
following trends:
\begin{enumerate}
\item The orbit of the innermost BH decays on the dynamical friction
  timescale.  As it does, the accretion rate goes from sub-Eddington
  Bondi-Hoyle-Littleton ($\dot{M}\propto M^2$, $\dot{M} \le
  \dot{M}_{\rm Edd}/\eta$) to Eddington-limited accretion
  ($\dot{M}=\dot{M}_{\rm Edd}/\eta\propto M$).  However, this phase
  lasts much less than a Salpeter time, and the mass growth is
  typically insignificant.
\item As the BH approaches the center of the halo, typically at $r\sim
  {\rm a~few}\times r_{\rm c}$, the Bondi rate $\dot{M}_{\rm
    B}>3000\dot{M}_{\rm Edd}$.  At this point, following
  \citetalias{Inayoshi+16}, we assume that photon trapping allows for
  hyper-Eddington accretion, i.e. once again matching the unimpeded
  BHL rate ($\dot{M}=\dot{M}_{\rm B}\propto M^2$).
\item In the hyper-Eddington phase, the BH grows from a typical mass
  of a few $\times 10\Msol$ to $\ga 10^3\Msol$.  During this rapid
  transition, the BH becomes more massive than the mass contained
  inside its orbit around the halo center ($M>M_{\rm bg}$), and the
  gas mass enclosed within its Bondi sphere exceeds its own mass
  ($M_{\rm B}>M$).
The behaviour up to this point is almost identical for all the
simulations that allow for hyper-Eddington accretion (models ``$f_{\rm
  in}=1$'',``$f_{\rm in}=10^{-3}$'',``$f_{\rm in}=0$'' and ``I'').
Most of the variation between these models result from the difference
in prescriptions when $M_{\rm B}>M$ and $M>M_{\rm bg}$.  That is, the
models ``branch out'' from this point forward.
\item The accretion rate then slows, as it encounters the halo mass
  supply limit $f_{\rm in}\dot{M}_{\rm in}$ or the asymptotic constant
  value of $\dot{M}_{\rm B}$ in the limit of large Bondi radius.
  Because both of these values are constant, as the BH mass increases
  $\dot{M}$ falls below $3000\dot{M}_{\rm Edd}\propto M$, and becomes
  Eddington-limited again.  The final masses of the BHs, and their
  configuration with respect to the halo and other BHs, depend on the
  prescriptions for accretion and background gravitational force, as
  discussed in the following.
\end{enumerate}

\subsection{The final IMBH masses and configurations}
\label{subsec:finalmasses}

In Table \ref{table:summary2}, we summarize the average final masses
and accretion rates found in our simulations, for each combination of
our prescriptions for gas accretion and treatment of the background
gravitational force.  The values presented are the mean values over
the 24 realizations per model with a nuclear BH, evaluated when the first BH
has sunk to the center of the protogalaxy and has captured a companion
BH into a closed orbit with a semimajor axis $\approx 1\pc$.
 We have chosen to stop the simulations at $a=1\pc$,
because at smaller separations between the innermost bound pair three-body
scatterings are rare. We assume that past this separation, the inner bound pairs
evolve through damped, closed orbits until merger. masses and instantaneous accretion rates of the central BH are denoted
with a subscript ``1,'' and those for the smaller companion BH with a
subscript ``2.''  We also list the mass ratio $M_2/M_1 \le 1$ and the
orbital eccentricity $e$ of the pair.

The bottom half of Table~\ref{table:summary2} lists the BH properties
found in simulations where the background gravitational force exerted
on a given BH was set to zero whenever the BH mass exceeded the mass
enclosed inside its radial position. The final values found in these
simulations do not vary significantly from the ones in which the
background force was always present (the top half of the table).

\subsubsection{The central BH}

Let us first discuss the central BH.  In the simulations with $f_{\rm
  in}=0$, BHs stop growing once they are outweighed by the gas mass
enclosed inside their Bondi sphere. Because of this, they never grow
beyond $M_1\sim 100 \Msol$.  In simulations where $f_{\rm
  in}=10^{-3}$, $f_{\rm in}=1$, and ``I'' the growth rate is capped
by $f_{\rm in}\dot{M}_{\rm in}$ and by the asymptotic Bondi rate
(eq.~\ref{Bondiasymptoticvalue}).  These upper bounds allow the growth
of the central BH to $M_1\sim 10^4\Msol$ for $f_{\rm in}=10^{-3}$, and
$M_1\sim 10^5\Msol$ to $\sim 10^6\Msol$ for $f_{\rm in}=1$ and ``I''
(note that the values for $f_{\rm in}\dot{M}_{\rm in}$ and the
asymptotic Bondi rate are comparable).  \textit{In all simulations
  where hyper-Eddington accretion is allowed to continue past the
  point $M<M_{\rm B}$, the central BH grows into an IMBH or SMBH.}

For reference, we can see that if the mass growth is limited at the
canonical Eddington value (model ``E''), then the central BH which
forms a bound pair with another BH does not grow significantly. This
is because the bound pairs reach $a\la 1\pc$ soon after the first BH
sinks to the core, when both BHs are still close to their original
stellar masses.  Once a tight central binary forms, its orbital
velocity increases, and suppresses the BHL accretion rate below the
Eddington value, stunting further growth of either BH.
We also show, for reference, values in which accretion is not allowed
at all (model ``$\dot{M}=0$'').

\subsubsection{The stellar-mass BH companion}

A striking result of our simulations is that we find no more than one
hyper-accreting BH per simulation (i.e. either zero or one IMBH per
galaxy).  The reason for this is that if an IMBH forms in one of our
simulations, it prevents other BHs from undergoing hyper-Eddington
accretion.

This finding can be explained as follows.  The first IMBH forms
relatively quickly (its orbit decays in a few Myr), and does so at the
center of the halo, where gas densities (and Bondi accretion rates)
are high.  Once this IMBH is in the center of the halo, any subsequent
BH whose orbit decays will be captured by the IMBH potential.  Whereas
the first BH had orbital velocities $v\la c_{\rm s}$ as it fell toward
the center, the orbital velocity of a BH captured by the IMBH will be
supersonic.  The supersonic orbital motion suppresses both the orbital
decay rate via dynamical friction, preventing the second BH from
sinking deep into the gas-rich center of the halo.  On top of this,
the high velocity suppresses the Bondi-Hoyle-Lyttleton accretion rate,
$\dot{M}_{\rm B}\propto (c_{\rm s}^2+v^2)^{-3}$.
   
Thus, the first BH to wander to the center of the potential is able to
grow to an IMBH via hyper-Eddington accretion, but then
\textit{subsequently prevents other BHs from doing the same}.  As a
result, our simulations typically produce bound pairs of IMBHs and
stellar-mass BHs with mass ratios $q\equiv M_1/M_2 \sim
10^{-4}-10^{-2}$.  Such ``extreme mass-ratio'' systems are one of the
important targets for detection by planned gravitational-wave
instruments, and we will revisit them in our discussion section.  As
Table~\ref{table:summary2} shows, interestingly, all EMRIs have a
highly eccentric orbit; this is a result of the preferential capture
of stellar-mass BHs on such orbits, i.e. with pericenters inside the
sphere of influence of the newly grown IMBH.

In models where an IMBH is not produced, subsequent BHs are free to
fall to the center at low speeds, just as the first one did.  As shown
in Table~\ref{table:summary2}, in the ``$f_{\rm in}=0$'' models, we
find near-equal, stellar-mass binaries forming in the nucleus.  In
these models, the growth of the 1st BH is artificially stunted,
allowing the 2nd BH to experience a brief phase of hyper-Eddington
accretion.  This hyper-Eddington phase, as the 2nd BH travels through
the dense gaseous core, only lasts until it, too, reaches a mass of
$M_2\ga 100\Msol$; its Bondi sphere will then become self-gravitating,
and its grow is terminated, just as for the first BH.

\vspace{-\baselineskip}

\subsubsection{Different treatments of central background potential}

The main difference in the two sets of models (shown in the bottom
vs. top half of Table~\ref{table:summary2}) is that if $a_{\rm bg}$ is
always present, the central BH ends up at the very center of the model
protogalaxy, because the background force always continues to point
inward.  In contrast, when $a_{\rm bg}$ is turned off, the gas drag
brings the BHs to rest near---but not \textit{at}---the center.  Aside
from this detail, we find no major qualitative difference in the
properties of the BHs across these two sets of simulations.  We
conclude that the hydrodynamical reaction of the innermost gas to the
central BH is unlikely to significantly influence our main conclusions
about the demography and location of the emerging BH population (in
particular whether hyper-Eddington accretion occurs).

\section{Discussion}
\label{sec:discuss}

\subsection{SMBH precursors}

Our simulations focus on the growth and orbital evolution of Pop~III
remnant BHs in a model protogalaxy that is just above the
atomic-cooling threshold for virial mass.  We find that the
hyper-Eddington accretion prescription of \citetalias{Inayoshi+16}
typically results in the formation of a single IMBH in the center of
the protogalaxy.

The natural interpretation is that this is a massive nuclear BH that
will continue to grow as the host galaxy grows.  The formation of a
$\sim (10^4-10^5) \Msol$ BH in an atomic-cooling halo is the same end
result as in the so-called ``direct collapse BH'' scenario.  These
IMBHs must then grow at a logarithmically time-averaged rate
$\dot{M}\la 10 \dot{M}_{\rm Edd}$ (i.e. at a rate comparable to the
Eddington limit for a radiative efficiency $\eta\sim 0.1$) to explain
the $\ga 10^9\Msol$ engines of the $z\ga 6$ quasars.  The direct
collapse scenarios generically require specific conditions that may be
rare in the Universe. For example, in most models the collapse is
facilitated by a high Lyman-Werner intensity that dissociates
molecular hydrogen, and thus only a small fraction of galaxies are
expected to be viable direct-collapse sites
(e.g. \citealt{Dijkstra+08, Shang+10, Hosokawa+12,
  Dijkstra+14,Sugimura+2014, Visbal+14, Latif+15,
  InayoshiTanaka15,Regan+2016}).

The picture suggested by our simulations is that IMBHs in
protogalactic nuclei could plausibly be more commonly produced by
hyper-Eddington growth of a pre-existing stellar-mass BH -- the
essential requirements being only a high-density core, and a
large-scale inflow rate of $O({\rm \Msol yr^{-1}})$, down to the Bondi
radius ($\sim$0.01 pc) of the central BH with initial mass of $\sim
100\Msol$.

These two hypotheses for SMBH progenitors---rare direct-collapse seeds
and more common results of hyper-Eddington accretion---could be tested
against observations through event rates detected by milli-Hertz
gravitational-wave detectors (e.g. \citealt{Sesana+07,dyformula1}) or
by the global signatures of the redshifted 21 cm line
(e.g. \citealt{TOP16}).

\subsection{EMRI detections}

Whenever an IMBH forms in our simulations, we find that it captures
one or more stellar-mass BH companions.  Mergers of such BHs are
predicted to produce EMRIs, a category of gravitational wave events
that is one of the primary \textit{low-redshift} targets of the
space-based interferometer \textit{eLISA} \citep{eLISAwhitepaper,
  eLISAGWN}.

Because the timescale for the $M_2\ll M_1$ pairs in our simulations to
merge through emission of gravitational waves is well over a Hubble
time, additional mechanisms such as three-body scatterings or
continuous gaseous dissipation (e.g. by a circumbinary accretion disc, \citealt{Cuadra+09,Roedig+11})
are required to drive the merger.
The production of such pairs in our simulations suggests that they could
result in mergers of IMBHs and stellar-mass BHs at lower redshifts.

We note that our IMBH-BH pairs have eccentricities $e\ga 0.9$, at
$a\approx 1 \pc$ (Table \ref{table:summary2}).  This points to the
interesting possibility that they could lead to EMRIs that have
residual eccentricities when they enter the eLISA band.  However, as
we do not follow the evolution of such pairs all the way to merger,
and given the variety of possibly relevant mechanisms, we leave the
assessment of any residual eccentricity in the eLISA band for future
work.

Additionally, the merger of a protogalaxy or dwarf galaxy containing
an IMBH with a more massive one containing a SMBH should result in the
formation of a SMBH-IMBH pair.  While this is an expected corollary of
our findings, further work is required to assess whether such pairs
can overcome the so-called ``final parsec problem'' \citep{MM05}.

\subsection{Comparison with previous work}

Several recent papers have investigated super-Eddington accretion
in galaxies.
\cite{Lupi+16} used hydrodynamical simulations to investigate
super-Eddington accretion of stellar-mass BHs in circumnuclear gas discs
in the hearts of galaxies. Their scheme allows gas particles close to the BH
to accrete and grow the BH.
\cite{Pezzulli+16} considered the growth of stellar-mass BHs inside model
galaxies that account for metal enrichment, dust, star formation and
detailed cooling. They assume that the central BH
grows at a rate proportional to the cold gas mass in the bulge,
and inversely proportional to the dynamical time of the bulge.

One major difference between this study and those papers
is the accretion prescription. We adopt the analytic Bondi-like accretion
prescription based on \citetalias{Inayoshi+16} and featuring
the transition from low Bondi-Hoyle-Lyttleton accretion,
to Eddington-limited accretion, and then to hyper-Eddington accretion.
Another notable difference is that \cite{Lupi+16} and \cite{Pezzulli+16} 
examined the growth of BHs in fully evolved galaxies,
whereas in this study we focus on the growth of Pop~III remnant BHs
into IMBHs in a protogalaxy of mass $\sim 10^8\Msol$.

\citet{AlexanderNatarajan2014} considered a set-up similar to ours, in
which a stellar-mass BH is in orbit in a protogalaxy, accreting above
the fiducial Eddington rate. The focus of that paper was to assess the
ability of the gas inside the BH's sphere of influence to shed angular
momentum and accrete onto the BH. The orbit of the BH was assumed to
be determined by its interactions with a nuclear star cluster (which
was found to be important to reduce the angular momentum). Here we
treated a small system of BHs, and assumed that the background gas
dominates their orbital decay into the nucleus.

Closest to our study, \citet{Tagawa+15} and \citet{Tagawa+16}
performed $N$-body simulations of BHs embedded in a compact
distribution of gas, in order to gauge the merger mechanisms of BHs in
galactic centers.  Overall, the set-up and goals of these studies and
ours are similar, although Tagawa et al.'s focus was to clarify the
occurrence rate and mechanisms of stellar-mass BH mergers.  Our most
notable finding -- the frequent formation of a \textit{single IMBH} at
the \textit{center} of the protogalaxy -- differs from the
conclusions by \citet{Tagawa+15} and \citet{Tagawa+16}, who find
efficient formation of stellar-mass binaries, often facilitated by
3-body interactions.

These differences in conclusions arise from three important
differences between our initial conditions and model assumptions.
First, we spread 10 initial BHs over a large region of up to $100$~pc,
with separations of $>10$~pc. The initial BH separations are much more
compact in \citet[][$0.01-10~$pc]{Tagawa+15} and especially in
\citet[][$0.01-0.1$~pc]{Tagawa+16}.  As a result, we do not find
3-body interactions or stellar-mass binaries.
Second, we adopt a centrally condensed density profile, while
\citet{Tagawa+15} and \citet{Tagawa+16} both assume homogeneous
clouds.
Third, we investigate various accretion prescriptions, and allow
hyper-Eddington accretion at rates limited only by the steady
large-scale inflow rate (whereas \cite{Tagawa+15} did not consider
accretion onto BHs and \cite{Tagawa+16} considered an accretion rate
capped by the Eddington limit and by the assumed total cloud mass).
As a result of these last two differences, we find that rapid growth
into IMBHs and SMBHs is much more common, and always occurs in the
nucleus -- producing EMRIs, rather than stellar-mass binaries.

\subsection{Caveats}

Our results were obtained in simplified toy models, and are subject to
several caveats.  We here discuss three possible major limitations of
our model.

\vspace{\baselineskip}
\noindent\textit{Gravitational potential of the background matter.}
Our simulations assume a static density profile of gas and dark
matter, instead of allowing the protogalaxy to evolve dynamically or
thermodynamically.  On one hand, the simplified treatment of the
protogalaxy allowed us to run hundreds of simulations---dozens of
initial condition realizations for each of a dozen different
combinations of theoretical models.  On the other, this survey of
model prescriptions and the statistical sample size came at the
expense of a more detailed treatment of gas dynamics.

In particular, not accounting for the dynamical evolution of the
surrounding matter directly impacts the two gravitational effects in
this work: the gravitational force exerted by the ambient matter, and
dynamical friction.  Our assumption of a static background allows us
to evaluate analytically the gravitational force from the ambient
medium, and we treat the dynamical friction using a modified version
of the Chandrasekhar formula \citep{Galacticdynamics}.

In an attempt to gauge the robustness of our results with respect to
these theoretical simplifications, we ran different sets of
simulations with very different assumptions.  First, as discussed
above, we ran a set of simulations in which the gravitational field of
the background material was entirely removed, once the BH mass
exceeded the mass of the matter enclosed inside its orbit.  The only
qualitative difference we found was that this resulted in the central
BH ending up slightly off-center (whereas leaving the analytic
background force ``on'' had the effect of always pulling the central
BH to the very center of our model protogalaxy). We also note that while gas anisotropies are common features in simulations
of protogalactic haloes, the masses of any anisotropic clumps
tend to be small compared to that of the ambient gas
(which are overall well-described by power-law profiles,
e.g. \citealt{Regan09}).

We also ran, for the same set of the different accretion
prescriptions, simulations in which the dynamical friction was not
calculated using the local gas density at the BH coordinates
$\rho(r)$, but using the gas density averaged over the surface of its
Bondi sphere, $\langle\rho_{\rm B}(r)\rangle$.  The rationale behind
this experiment was that whereas the derivation of the Chandrasekhar
formula assumes an infinite, uniform background distribution of gas,
the physical phenomenon of dynamical friction is due to the wake of
overdense gas formed at some distance from the massive body.  We found
no qualitative difference between this set of simulations and the one
described in \S3.  While there is a rich literature on quantifying how
dynamical friction differs in nonuniform density distributions or
nonlinear trajectories
\citep[e.g][]{Brandenburg01,Just05,Kim07,Kim10}, these studies do not
report major differences from the Chandrasekhar formula.

We conclude that our results are unlikely to be an artifact of the
simplified treatment of the gravity of the ambient matter
distribution.

\vspace{\baselineskip}
\noindent\textit{Star formation in the halo.}
Our model does not consider star formation inside the halo.
 Once a dense region of gas cloud in the halo  becomes optically thick,
 fragmentation leads to star formation  \citep[e.g.][and refs. therein]{Regan+14,Becerra+15}.
 The stars may form directly in the core region,
 or form elsewhere and subsequently migrate to the core.
 Because the stars have lifetimes of $\sim\Myr$, we expect them to
 become BHs before arriving at the core, or shortly afterward.
 In practice, we do not expect significant qualitative differences
 based on which type of BH reaches the core first---the pre-existing BHs
 our simulations had in mind, or the stars/BHs that form in situ in the halo.
 According to the picture suggested by our simulations,
 whichever type of BH arrives at the center first should grow massive via
 hyper-Eddington accretion, then capture any subsequent
 arrivals into bound orbits (which may eventually produce EMRI events).
 
Therefore, we do not expect any in situ star formation in
 the halo to qualitatively affect our findings.

\vspace{\baselineskip}
\noindent\textit{Effect of radiation on the hydrodynamics.}  Our toy
model neglects the radiation produced by the accreting BHs (and of any
stars found in the same galaxy).  The gas in the halo cools
efficiently and is mostly neutral, and accreting BHs will create their
individual small HII regions.  Here we estimate the size of these HII
regions, before the BHs wander into the core and reach the
hyper-Eddington state.  In the ``intermediate'' regime, when
$\eta^{-1} \la \dot{M}_{\rm B}/\dot{M}_{\rm Edd}\la 3000$, the
time-averaged accretion rate is limited to the Eddington rate, because
of the periodic formation, disappearance, and re-appearance, of an HII
region that makes the accretion episodic.  The maximum size of this
HII region is larger than the Bondi radius, by definition, before the
hyper-Eddington state can be reached \citepalias{Inayoshi+16}.
Assuming a luminosity of $L_{\rm Edd}$, the HII region size is $R_{\rm
  HII} = 8\times 10^{13} (M/30\Msol)^{1/3} (r/r_c)^{4/3}{\rm
  cm}$ (see eq.~27 in \citetalias{Inayoshi+16}) or $R_{\rm HII}/r =
0.01 (M/30\Msol)^{1/3} (r/r_c)^{1/3}$.  This means that the
HII region remains relatively small for BHs located within a few pc of
the core. A near-Eddington BH outside this region (say at 10-100~pc)
could blow a large HII bubble and change the global density
distribution.  However, the stellar-mass BHs in these outer,
low-density regions will be highly sub-Eddington and dim
(eq.~\ref{eq:mdotbondi}).  We conclude that radiative feedback is
unlikely to prevent the onset of the hyper-Eddington phase of the
first BH that sinks to the central region.

\vspace{\baselineskip}
\noindent\textit{Validity of Hyper-Eddington accretion.}  As
emphasized throughout this paper, a key ingredient of our model is
that we allow rapid accretion, well in excess of $L_{\rm Edd}/c^2$.
This is based on the recent results of \citetalias{Inayoshi+16}, who
find this to be the case for accreting BHs whose HII region is
confined to within the Bondi radius.  This conclusion is subject to a
few caveats summarized in \citetalias{Inayoshi+16}.  In particular,
here we highlight the fact that \citetalias{Inayoshi+16} assumes a
spherically symmetric accretion flow with low angular momentum, such
that the centrifugal radius (setting the size of an accretion disc,
producing significant luminosity) is smaller than the trapping radius
(inside which photons are advected inward with the flow, rather than
diffusing out).  At the onset of the hyper-Eddington phase, the latter
is $1500~R_{\rm Sch}$, or $5\times10^{10}(M/100\Msol)$~cm.  We thus
require that the accretion flow onto the BHs remain quasi-spherical
down to this distance from the BH.  While this appears feasible once
the BH settles to the bottom of the potential well, a proper
assessment will require follow-up investigations, resolving the
angular momentum transfer and dissipation for the flow onto the
central BH.  However, we note here that this requirement is much less
stringent than the one addressed in a similar context by
\citet{AlexanderNatarajan2014}, who required the centrifugal radius to
be as small as a few $\times R_{\rm Sch}$ (and found it to be
feasible, faciliated in their model by resonant effects due to a
central stellar cluster).

\vspace{-2\baselineskip}
\section{Summary}

In this paper, we described the formation of $10^{3-5}\Msol$ IMBHs in
centrally condensed gas clouds, arising from a small cluster of
Population III remnant black holes (BHs).
The stellar-mass BH are
assumed to have been delivered into the cloud during the process of
the hierarchical assembly of the halo via mergers, and have a
spatially extended initial configuration (several pc to 100 pc). We
then follow their accretion and orbital dynamics via an N-body
calculation.  These calculations reveal that as a result of gas drag,
one of these BHs typically sinks to the nucleus, where it rapidly
grows into an IMBH.

Our results suggest a viable pathway to forming the earliest massive
BHs in the centers of early galaxies.  We also find that only one IMBH
can form in this way per galaxy, and that this IMBH typically captures
a stellar-mass BH companion, making these systems observable in
gravitational waves as extreme mass-ratio inspirals (EMRIs) with
\textit{eLISA}.

More detailed simulations that account for the hydrodynamics,
radiative transfer, and the cosmological evolution of the host
protogalaxy are required to test this idea further.

\vspace{0.5cm}

{\bf Acknowledgements} Results in this paper were obtained using the
high-performance LIred computing system at the Institute for Advanced
Computational Science at Stony Brook University, which was obtained
through the Empire State Development grant NYS \#28451.  This work was
supported in part by NASA grants NNX11AE05G and NNX15AB19G (ZH).

\end{document}